\documentclass[aps,reprint,superscriptaddress,showpacs]{revtex4-2} 
\usepackage{graphicx}
\graphicspath{{./figs/}}
\usepackage{amssymb}
\usepackage{amsmath}
\usepackage{natbib}
\usepackage{color}
\usepackage{hyperref}

\usepackage[textwidth=7.35in,top=1in,bottom=1in]{geometry} % to fit figures

\newcommand{\SM}{Supplementary}
\newcommand{\reffig}[2][]{Fig.~\ref{#2}\MakeLowercase{#1}}
\newcommand{\refeq}[1]{Eq.~(\ref{#1})}
\newcommand{\refsec}[1]{Sec.~\ref{#1}}

\begin{document}

\title{Microwave radiometry of a quantum-critical, hybrid Josephson array}
% \title{Critical radiation of the anomalous metal}
% \title{non-equilibrium radiation at quantum criticality probed in a Josephson junction array}

\author{Kristen W. L\'eonard}
\thanks{Equal contribution}
\affiliation{Institute of Science and Technology Austria, Am Campus 1, Klosterneuburg, 3400, Austria}

\author{Anton V. Bubis}
\thanks{Equal contribution}
\affiliation{Institute of Science and Technology Austria, Am Campus 1, Klosterneuburg, 3400, Austria}

\author{Melissa Mikalsen}
% \email{}
\affiliation{Department of Physics, Center for Quantum Phenomena, New York University, New York, NY, 10003, USA}

\author{William F. Schiela}
% \email{}
\affiliation{Department of Physics, Center for Quantum Phenomena, New York University, New York, NY, 10003, USA}

\author{Bassel H. Elfeky}
% \email{}
\affiliation{Department of Physics, Center for Quantum Phenomena, New York University, New York, NY, 10003, USA}

\author{William M. Strickland}
\affiliation{Department of Physics, Center for Quantum Phenomena, New York University, New York, NY, 10003, USA}

\author{Duc Phan}
% \email{}
\affiliation{Institute of Science and Technology Austria, Am Campus 1, Klosterneuburg, 3400, Austria}

\author{Javad Shabani}
% \email{}
\affiliation{Department of Physics, Center for Quantum Phenomena, New York University, New York, NY, 10003, USA}

\author{Andrew P. Higginbotham}
\email{ahigginbotham@uchicago.edu}
\affiliation{James Franck Institute and Department of Physics, University of Chicago,  929 E 57th St, Chicago, Illinois 60637, USA}
\affiliation{Institute of Science and Technology Austria, Am Campus 1, Klosterneuburg, 3400, Austria}

\begin{abstract}
Arrays of Josephson junctions can be tuned through anomalous metallic, quantum-critical, and insulating regimes.
% If persistent to zero temperature, anomalous metallic behavior poses a challenge to theory.
% Here, we show that incomplete thermalization between the sample and the cryostat is a cause of this behavior.
We introduce a new experimental probe, capturing microwave radiation across all three regimes, using a two-dimensional array of superconductor-semiconductor hybrid Josephson junctions as a model system.
Our approach allows \textit{in-situ} calibration of the sample's circuit parameters and provides isolation from measurement back-action effects.
We measure the radiation temperature of the anomalous metal, and find that it is hotter than both the quantum-critical and insulating regimes.
We further show that the anomalous-metallic regime is more susceptible to additional heating than other regimes, explaining its emergence in otherwise thermalized systems.
Turning to the quantum-critical regime, we discover nonlinear scaling of radiative noise with applied bias, consistent with theoretical predictions of universal non-equilibrium behavior at quantum critical points.
\end{abstract}

\maketitle

\section{Introduction}

In weak superconductors, quantum fluctuations lead to a zero-temperature quantum phase transition from superconducting to insulating behavior \cite{sacepe_quantum_2020}.
On the superconducting side of the transition, anomalous metallic resistance saturation is commonly observed at low temperature ~\cite{jaeger_onset_1989,ephron_observation_1996,crauste_thickness_2009,eley_approaching_2012,han_collapse_2014,boettcher_superconducting_2018,boettcher_dynamical_2023,ienaga_broadened_2024}.
% The anomalous metallic regime occurs in a strikingly wide array of platforms and materials, including the Josephson-junction arrays under study in this work.
Anomalous metallic behavior challenges theoretical understanding by suggesting a two-dimensional metal at $T=0$~\cite{abrahams_scaling_1979,kapitulnik_colloquium_2019}, and has characteristics of a distinct thermodynamic phase~\cite{boettcher_Berezinskii_2022}.
On the other hand, in several cases it has been experimentally shown that increased filtering eliminates or reduces anomalous-metallic behavior, indicating that the phenomenon could be due to a lack of thermalization~\cite{tamir_sensitivity_2019,shin_effect_2020}.
Recently, the low-frequency noise of an anomalous-metallic system was measured, confirming the presence of non-equilibrium behavior \cite{haug_excess_2024}.
Our work adds to this effort by finding non-equilibrium behavior in a second, established model system, while providing novel features such as one thousand times higher measurement frequency, \textit{in-situ} calibration of the sample's circuit parameters, and full isolation from undesirable back-action effects.
% Currently, it is unclear whether anomalous metallic behavior reflects a lack of thermal equilibrium with the cryostat, crossover behavior, or a distinct thermodynamic phase of matter.
% We will show that this question can be resolved by measuring radiation emitted in the anomalous metallic regime.

A second and equally important motivation for our work is to test predictions of universal non-equilibrium behavior near quantum critical points.
Analyses based on critical field theories~\cite{dalidovich_nonlinear_2004,green_nonlinear_2005,green_current_2006} and gauge-gravity duality~\cite{sonner_hawking_2012} have predicted that, near the superconductor-insulator quantum phase transition, non-equilibrium noise is described by an effective temperature whose scaling with applied bias is controlled by the dynamical critical exponent.
Our work tests this theoretical prediction.

Both thermometry of the anomalous metal and testing predictions of non-equilibrium behavior near criticality require a non-invasive and calibrated probe of sample temperature.
% Testing these predictions requires an experimental apparatus that can of non-equilibrium behavior, moving beyond the traditional transport probes of superconductor-insulator physics.
Here, we meet these requirements by measuring microwave radiation, which has three experimental advantages.
First, well-established methodology from the calibration of microwave radiometers~\cite{dicke_atmospheric_1946}, quantum-limited amplifiers~\cite{castellanos-beltran_amplification_2008}, and axion haloscopes~\cite{brubaker_first_2017}, allows an accurate measurement of the temperature of radiation emitted by the sample.
Second, the availability of circulators at our gigahertz measurement frequencies allows for nearly ideal isolation of the fragile sample from the measurement apparatus.
Third, microwave scattering parameters can be measured \textit{in-situ}, allowing us to convert emitted microwave radiation power into an equivalent sample temperature.

We use this new experimental approach to perform thermometry of the anomalous metal and study non-equilibrium scaling near the superconductor-insulator phase transition, focusing on an array of superconductor-semiconductor hybrid Josephson junctions as a tunable model system~\cite{eley_approaching_2012,boettcher_superconducting_2018}.
We observe excess radiation coincident with the onset of resistance saturation.
Converting the excess radiation into an equivalent sample temperature, the extracted temperature is found to be compatible with the sample falling out of equilibrium with the cryostat in the anomalous metallic regime.
We further find that the anomalous metallic regime is more susceptible to heating than either the quantum critical or insulating regimes.
Motivated by predictions of universal scaling behavior near quantum critical points~\cite{dalidovich_nonlinear_2004,green_nonlinear_2005,green_current_2006,sonner_hawking_2012}, we then move on to a study of radiation in the quantum-critical regime.
Establishing a non-equilibrium steady state using an applied bias, we find that the noise-equivalent radiation temperature is compatible with the theoretically predicted $\sqrt{I}$ scaling as a function of bias $I$.
% Motivated by predictions of universal scaling behavior near quantum critical points~\cite{dalidovich_nonlinear_2004,green_nonlinear_2005,green_current_2006,sonner_hawking_2012}, w
We observe a surprising collapse of non-equilibrium noise curves near the critical point, suggestive a broad regime with a unified non-equilibrium description near quantum criticality.
Adding to a recent flurry of experimental activity on non-equilibrium properties of quantum-critical systems such as ultraclean graphene \cite{majumdar_universality_2025} and heavy fermion strange metals \cite{chen_shot_2023}, these observations open an exciting frontier for exploration of non-equilibrium behavior of critical systems.
% Quantitatively however, our results deviate from the predictions in several ways, opening an exciting new frontier for exploration of non-equilibrium behavior of critical systems.
% With our apparatus we thereby demonstrate a new experimental path for studying the non-equilibrium response of quantum critical systems.

\begin{figure}
    \centering
    \includegraphics[width=3.54in]{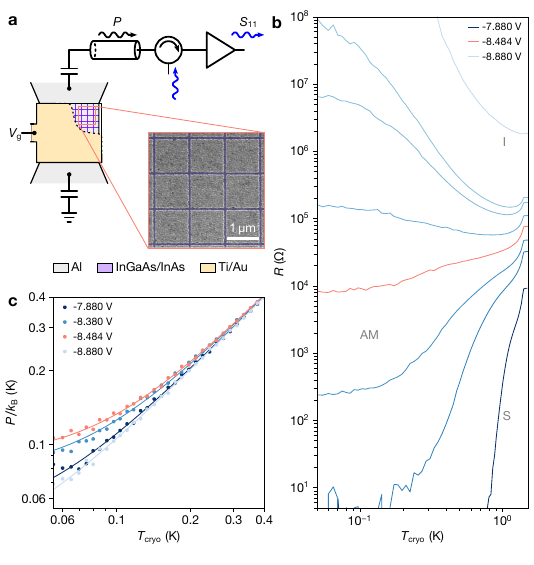}
    \caption{\textbf{a},~Schematic of the device showing the Al/InAs Josephson junction array capacitively coupled to a microwave readout chain, with a top-gate to tune carrier density in the InAs two dimensional electron gas.
    A circulator and filtering (not pictured) prevents back-action noise from disrupting the anomalous metallic state, and allows it to be probed with a weak microwave tone.
    The incident noise power spectral density $P$ can also be measured.
    Standard four-probe transport measurement is omitted from the schematic for clarity.
    For details of the measurement setup, see \SM~Figs.~\ref{fig:fig_fridge}-\ref{fig:fig_fridge_dc}.
    Scanning electron micrograph of the Al islands is shown in the inset. 
    \textbf{b},~Measured resistance $R$ versus cryostat temperature $T_\mathrm{cryo}$ at different top-gate voltages (indicated by colors). 
    \textbf{c},~Measured radiation $P$ as a function of $T_\mathrm{cryo}$ for selected top-gate voltages. 
    The most excess radiation is observed at a top-gate voltage of $-8.484~\mathrm{V}$, shown in panels~\textbf{b} and~\textbf{c} in salmon color. 
    Solid lines are a model for an effective saturation radiation, $P_\mathrm{sat}$, determined via best fit to each curve. 
    The values of $P_\mathrm{sat}/k_B$ for each top-gate voltage are $-7.88~\mathrm{V}$: $49~\mathrm{mK}$; $-8.38~\mathrm{V}$: $76~\mathrm{mK}$; $-8.484~\mathrm{V}$: $89~\mathrm{mK}$; $-8.88~\mathrm{V}$: $37~\mathrm{mK}$.}
    \label{fig:fig1}
\end{figure}

\section{Excess noise near the superconductor-insulator transition}

The key technical feature of our experiment is to connect an anomalous metallic system to a microwave readout chain, allowing us to accurately measure the radiation emitted from our sample.
That radiation (\reffig[a]{fig:fig1}, black arrow) is efficiently collected via coupling capacitors, while a circulator and filtering prevent measurement back-action effects on the fragile anomalous metallic state.
The anomalous metal is realized in a two-dimensional array of superconductor/semiconductor Al/InAs Josephson junctions~\cite{shabani_two-dimensional_2016,mayer_superconducting_2019}, formed from the two-dimensional electron gas situated in the $4~\mathrm{nm}$-thick InAs quantum well located $10~\mathrm{nm}$ below the Al layer.
Measurements of two devices (S1, S2) are presented, with all reported measurements from device S1 unless otherwise indicated.
Device S1 consists of a $40~\text{wide}\times100$~long array of Al squares with periodicity $1.15~\mathrm{\mu m}$, spaced by $0.08~\mathrm{\mu m}$ (\reffig[a]{fig:fig1}), and device S2 has the same proportions as S1 but the opposite aspect ratio: a $100~\text{wide}\times40$~long array of Al islands. 
In the superconducting regime these junctions are typically ballistic~\cite{mayer_superconducting_2019}.
In both devices a $50~\mathrm{nm}$-thick Al top-gate is deposited over a $60~\mathrm{nm}$ Al$_2$O$_3$ dielectric layer covering the entire array. 
The measurements were carried out in a dilution refrigerator with a base temperature of $10~\mathrm{mK}$.
Noise measurements carried out in a $10~\mathrm{MHz}$ band around $1.42~\mathrm{GHz}$ are reported as an averaged power spectral density, $P$, and expressed in temperature units as $P / k_\mathrm{B}$.
Uncalibrated microwave reflection ($S_{11}$, blue arrows in \reffig[a]{fig:fig1}) was measured and averaged in the same frequency band.
Microwave spectroscopy measurements were carried out in a wider band of $1.15-1.75~\mathrm{GHz}$ defined by the cryogenic circulator.
Complementary measurements at $5.2~\mathrm{GHz}$ in a $200~\mathrm{MHz}$ band are presented in \SM~\ref{sec:additional_sample}, with similar results observed.
Standard four-probe resistance measurements were performed with an AC bias of $5~\mathrm{\mu V}$ over the lines in the cryostat and the sample. 
All resistance values are reported as absolute resistance, except where sheet resistance is explicitly specified.
A detailed circuit diagram of the measurement apparatus is provided in \SM~\reffig{fig:fig_fridge}.

It is useful to categorize sample behavior based on the dependence of measured resistance on cryostat temperature.
For modest top-gate voltages, resistance decreases sharply with temperature, falling below the measurement threshold of our equipment, indicating a superconducting state (S, \reffig[b]{fig:fig1}).
In contrast, for more negative top-gate voltages resistance increases sharply with temperature, indicating insulating behavior (I, \reffig[b]{fig:fig1}).
At intermediate top-gate voltages there is an initial drop in resistance as the cryostat temperature is lowered, followed by a low-temperature saturation to a small but nonzero value (AM, \reffig[b]{fig:fig1}).
This low-temperature saturation near the critical point of the superconductor-insulator transition (SIT) is the hallmark behavior of the anomalous metal.

The observed superconducting, anomalous metallic, and insulating regimes qualitatively reproduce previous work on the same platform~\cite{boettcher_superconducting_2018,sasmal_voltage-tuned_2025}.
Although the resistance saturation we observe in the superconducting and insulating regimes is more symmetric than in Ref.~\cite{boettcher_superconducting_2018} it is similar to Ref.~\cite{sasmal_voltage-tuned_2025}, and to observations in other systems \cite{jaeger_onset_1989,kapitulnik_colloquium_2019}.
As we show later (Fig.~\ref{fig:addednoise}) weak external drives can dramatically increase saturation on the superconducting side while only weakly altering saturation on the insulating side, allowing us to convert our device to a strongly asymmetric case \textit{in-situ}.
We have also observed strong low-temperature saturation on the superconducting side in different measurement configurations (see \SM~\reffig{fig:sample0} and \SM~\reffig{fig:sample0_highfreq}), which later reduced after improvements to the experimental setup.
% Following Ref.~\cite{kapitulnik_colloquium_2019} we focus our discussion on saturation observed on the superconducting side of the transition.
Finally, we note that we have verified that resistance saturation occurs in samples that are not connected to microwave circuitry.

The simultaneously measured noise power spectral density $P$, referred to the input of the measurement chain, depends linearly on cryostat temperature for high $T_\mathrm{cryo}$, as expected based on the thermal equilibrium relation $P = k_B T_\mathrm{cryo}$ (\reffig[c]{fig:fig1}).
At lower temperatures $P$ saturates, indicating non-equilibrium behavior.
The data are well described by a phenomenological equation with a  gate voltage dependent saturation power $P_\mathrm{sat}$, according to $P^2 = k_B^2 T_\mathrm{cryo}^2 + P_\mathrm{sat}^2$.
The saturation power is largest when the sample is tuned near the critical resistance of the superconductor-insulator transition.
Studying the excess, non-equilibrium noise near the SIT is a central focus of this work.
In the following, we demonstrate the origin of the excess noise and show how to interpret it in terms of an effective sample temperature.

\section{Thermometry of the superconductor-insulator transition}

A more detailed understanding of excess noise at the superconductor-insulator transition can be obtained by studying the system's evolution with gate voltage at fixed cryostat temperature.
Measured device resistance increases with decreasing gate voltage (\reffig[a]{fig:fig2}).
At lower temperatures, the increase in resistance becomes more abrupt, with isothermal resistance curves crossing at a resistance separatrix of approximately $60~\mathrm{k \Omega}$, corresponding to a critical sheet resistance of $24~\mathrm{k \Omega}$.
For comparison, the (2+1)D $XY$ model has a critical sheet resistance of $22.6~\mathrm{k \Omega}$ in the clean limit~\cite{cha_universal_1991}, and a smaller value of $12.9~\mathrm{k \Omega}$ in the disordered case~\cite{swanson_dynamical_2014}.
Experimentally, critical points in other anomalous metallic systems have been reported at or near $h/4e^2 = 6.5~\mathrm{k \Omega}$~\cite{boettcher_superconducting_2018,haviland_onset_1989}.

For elevated cryostat temperature, negligible excess noise is observed at all gate voltages, a signature expected for a system in equilibrium (\reffig[b]{fig:fig2}).
However, at base temperature significant excess noise is observed, peaked slightly on the superconducting side of the SIT.
Elevated radiation levels are observed throughout the anomalous metallic regime identified in \reffig[b]{fig:fig1}.
These observations are compatible with the temperature-dependent noise study in \reffig[c]{fig:fig1}, and together indicate a broad region of non-equilibrium behavior surrounding the superconductor-insulator transition.
A sharp, solitary spike in excess noise is observed at $-8.16~\mathrm{V}$, well within the ostensibly superconducting region, possibly due to a collective plasma mode of the Josephson array.

\begin{figure}
    \centering
    \includegraphics[width=3.54in]{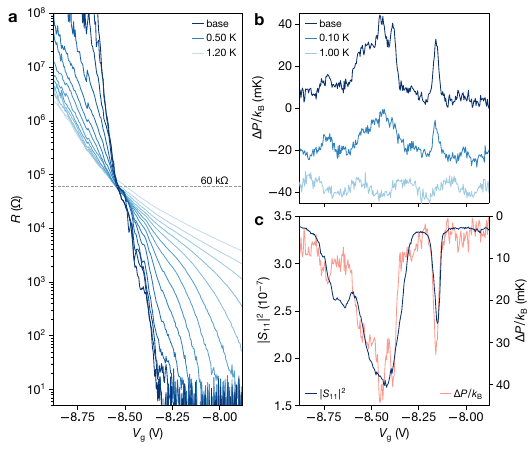}
    \caption{\textbf{a},~Measured resistance $R$ as a function of top-gate voltage $V_\mathrm{g}$ for different cryostat temperatures (indicated by colors).
    Resistance curves cross at the critical point with resistance $\approx 60~\mathrm{k \Omega}$ (indicated).
    \textbf{b},~Excess noise $\Delta P$ versus $V_\mathrm{g}$ for different cryostat temperatures.
    $\Delta P$ is defined as the measured noise power spectral density $P$ minus the value measured deep in the superconducting phase for each trace; it represents the emitted radiation in excess of thermal equilibrium.
    Large excess noise is observed at low temperature (darkest blue trace).
    Curves at $0.1~\mathrm{K}$ and $1~\mathrm{K}$ are shifted down for clarity by $20$ and $40~\mathrm{mK}$, respectively.
    \textbf{c},~Reflection coefficient $|S_{11}|^2$ (left axis) and $\Delta P$ at $10~\mathrm{mK}$ from panel \textbf{b} (right axis) versus $V_\mathrm{g}$.
    Excess noise is strongly negatively correlated with $|S_{11}|^2$.}
    \label{fig:fig2}
\end{figure}

Interestingly, excess noise is negatively correlated with the microwave reflection coefficient $|S_{11}|^2$ (\reffig[c]{fig:fig2}).
Qualitatively, this correlation reflects the fact that noise generated by the sample can only be efficiently measured when the sample is impedance matched, $S_{11}=0$.
Note that $S_{11}$ is weakly temperature dependent below $0.5~\mathrm{K}$ (see \SM~\reffig{fig:s11at500}), so that for a fixed gate voltage the temperature dependence in \reffig[b]{fig:fig2} should be attributed to a change in radiation, rather than a change in matching.
More quantitatively, the correlation between $S_{11}$ and excess noise is suggestive of a simple picture where the sample has some temperature $T_\mathrm{s}$ which is generally not equal to the cryostat temperature $T_\mathrm{cryo}$ (see Sec.~\ref{sec:circuit_theory}), and measured radiation is effectively filtered by an impedance mismatch factor.
This model predicts a linear relationship between $P$ and $|S_{11}|^2$,
\begin{equation}
    \label{eq:beamsplitter}
    P/k_B = -\alpha |S_{11}|^2~(T_\mathrm{s} - T_\mathrm{cryo}) + T_\mathrm{s},
\end{equation}
where $\alpha$ accounts for the net loss of the input lines and measurement chain, $\alpha |S_{11}|^2$ is the sample-referred reflection coefficient, and $\alpha |S_{11}|^2=1$ corresponds to complete reflection.
A similar model was used in~\cite{dieleman_observation_1997}.
Although $S_{11}$ and $\alpha$ are generally frequency dependent, we measure in a narrow band such that $S_{11}$ is constant, with negligible frequency dependence in $\alpha$ expected.

\begin{figure}
    \centering
    \includegraphics[width=3.54in]{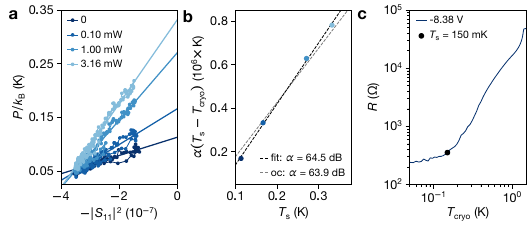}
    \caption{\textbf{a},~Parametric plot of noise power spectral density $P$ vs microwave reflection $|S_{11}|^2$, both measured as a function of gate voltage. 
    Power of microwave drive tone (at $547~\mathrm{MHz}$) at room temperature is indicated in legend; there is nominally $91~\mathrm{dB}$ of attenuation between room temperature and the sample.
    A linear fit is performed to each curve to extract $T_\mathrm{s}$ and $\alpha (T_\mathrm{s} - T_\mathrm{cryo})$ according to \refeq{eq:beamsplitter}.
    \textbf{b},~Slopes vs intercepts extracted from the fits shown in panel~\textbf{a}.
    A subsequent linear fit (black) yields the net loss $\alpha = 64.5~\mathrm{dB}$, and the cryostat temperature $T_\mathrm{cryo} = 50~\mathrm{mK}$. 
    The fit from the independent open circuit calibration (`oc', gray) of $63.9~\mathrm{dB}$ is also shown.
    Uncertainty is less than the marker size. 
    \textbf{c},~Measured resistance as a function of cryostat temperature is shown for the trace at $-8.38~\mathrm{V}$ from \reffig[b]{fig:fig1}, displaying anomalous metallic resistance saturation.
    $T_\mathrm{s}$ as determined from \reffig{eq:beamsplitter} using $\alpha$, $T_\mathrm{cryo}$ from the fits is marked on the resistance curve.
    Resistance saturation begins approximately at this temperature.}
    \label{fig:fig3}
\end{figure}

\begin{figure}[!ht]
    \centering\includegraphics[width=3.54in]{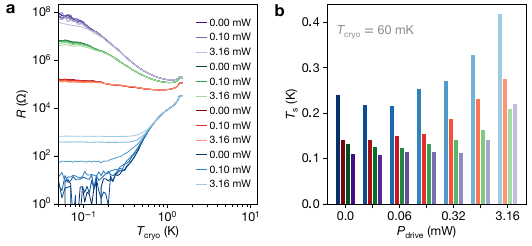}
    \caption{\textbf{a},~Sample resistance as a function of cryostat temperature at different gate voltages for various drive tone powers at $200~\mathrm{MHz}$. 
    Shown are superconducting/weakly anonomalous-metallic (blue, $V_g=-8.292~\mathrm{V}$),
    quantum-critical (red, $V_g=-8.552~\mathrm{V}$),
    and insulating (green, $V_g=-8.632~\mathrm{V}$ and purple, $V_g=-8.664~\mathrm{V}$).
    % Gate voltages are blue: -8.292~V, red: -8.552~V, green: -8.632~V, and purple: -8.664~V.
    % The resistance saturation associated with the anomalous metal increases with increasing drive tone power for applied powers between 0 and 3.16~mW.
    % The sample can be driven from a base temperature superconducting state to an anomalous metallic state with a resistance saturation onset temperature of up to approximately $400~\mathrm{mK}$.
    Resistance saturation is more pronounced on the superconducting side of the transition compared to the insulating side.
    \textbf{b},~Sample temperature $T_\mathrm{s}$ versus drive power $P_\mathrm{drive}$ for different gate voltages.
    Colors indicate gate voltage and shades indicate powers, matching \textbf{a}.
    Heating by $P_\mathrm{drive}$ is more pronounced on the superconducting side of the transition compared to the insulating side.
    }
    \label{fig:addednoise}
\end{figure}

To explore the noise model described by \refeq{eq:beamsplitter}, we intentionally increase $T_\mathrm{s}$ by driving the sample out of equilibrium with a second microwave tone far from our measurement band, measuring $P$ and $S_{11}$ for different drive strengths. %f=547~MHz.
We first focus on the coarse changes in $T_\mathrm{s}$ at each drive power (\reffig{fig:fig3}), and then go on to study the detailed evolution of $T_\mathrm{s}$ with both drive power and gate voltage (\reffig{fig:addednoise}).
At fixed driving power, $P$ and $|S_{11}|^2$ are approximately linearly related (\reffig[a]{fig:fig3}).
% , as expected for a gate-voltage independent $T_\mathrm{s}$, consistent with the previously discussed correlations between noise power and microwave reflection (\reffig[c]{fig:fig2}).
% As the drive power is gradually increased, $P$ and $|S_{11}|^2$ continue to maintain an approximate linear relationship.
Both the slope and vertical intercept increase with microwave drive power, while the intersection between curves is approximately drive-power independent.
These observations have a straightforward interpretation within the noise model~(\ref{eq:beamsplitter}).
Both increasing slope and vertical intercept reflect an overall change in $T_\mathrm{s}$ for each drive power.
The fixed crossing point between each line at  $-|S_{11}|^2=1/\alpha$ and $P/k_B=T_\mathrm{cryo}$ reflects a constant $T_\mathrm{cryo}$.
% Both increasing vertical intercept and slope can be used to extract a best-fit sample overall sample temeperature $T_\mathrm{s}$ at each drive power, and a fixed crossing point at

More formally, the best-fit slopes and vertical intercepts from \reffig[a]{fig:fig3} can be analyzed to extract $\alpha$ and the cryostat temperature $T_\mathrm{cryo}$, as shown in \reffig[b]{fig:fig3}.
The fitted value of $\alpha = 64.5~\mathrm{dB}$ is consistent with an independent open-circuit calibration, and the extracted value of $T_\mathrm{cryo} = 50~\mathrm{mK}$ was used throughout the analysis to calibrate out the added noise at the base temperature of the cryostat (see \SM~\refsec{sec:calibration})~\footnote{Since the determination of $P$ (e.g., in \reffig[a]{fig:fig3}) from the measured data requires knowledge of $T_\mathrm{cryo}$, this analysis enables a self-consistent extraction of $T_\mathrm{cryo}$. Specifically, the value of $T_\mathrm{cryo}$ obtained from the fits in \reffig[b]{fig:fig3} should match the initially assumed $T_\mathrm{cryo}$ used in the calculation of $P$ in \reffig[a]{fig:fig3}.}.
As a point of comparison, fixing $\alpha$ to the open-circuit calibration value of $63.9~\mathrm{dB}$ yields a best-fit $T_\mathrm{cryo}$ of $28~\mathrm{mK}$.
Note that inferred $T_\mathrm{cryo}$ values are in good agreement with the phenomenological saturation temperatures extracted in \reffig[c]{fig:fig1} for $V_\mathrm{g} = -7.88$\,V and $-8.88$\,V.
Indeed for these gate voltages $\alpha |S_{11}|^2 \approx 0.9$, so 90\% of the radiation seen by the amplifier comes from the $50\,\Omega$ termination of the circulator, thus allowing us to probe the degree of thermalization in the cryostat directly.
% The linear relationship between $P$ and $|S_{11}|^2$ and the compatibility of $\alpha$ and $T_\mathrm{cryo}$ with independent measurements gives evidence that \refeq{eq:beamsplitter} constitutes a predictive noise model for the device under study.

Having now experimentally calibrated the parameters $\alpha$ and $T_\mathrm{cryo}$ in \refeq{eq:beamsplitter}, it is possible to make an inference of sample temperature from a measurement of noise power spectral density.
In the case of the undriven anomalous metallic curve from \reffig[b]{fig:fig1}, we find $T_\mathrm{s}=150~\mathrm{mK}$, well above the base cryostat temperature determined either by the fits in \reffig[a]{fig:fig3}, by the open circuit calibration, or by readings of the RuO$_\mathrm{x}$ thermometer installed on the mixing chamber plate.
Rather, the extracted temperature of the anomalous metal coincides with the onset of anomalous metallic behavior, as shown in \reffig[c]{fig:fig3}.
Thus, the anomalous metallic resistance saturation can be naturally understood as a consequence of the sample falling out of equilibrium with the cryostat below $\approx 150~\mathrm{mK}$; it does not indicate a saturation in $R(T)$ as $T \rightarrow 0$.

For a more detailed exploration of the thermalization in different regimes, we now compare the extracted temperature of the anomalous metal with the critical and insulating regimes, both with and without intentional driving.
% Having performed thermometry of the anomalous metal, we now turn to a broader study of behavior near the SIT.
Figure~\ref{fig:addednoise}(a) shows representative $R(T)$ curves on the superconducting side of the anomalous metallic, quantum-critical, and insulating regimes for different applied microwave drive powers.
% Applying a weak microwave drive causes a dramatic increase in resistance saturation in the superconducting to anomalous metallic regime.
Applying a weak microwave drive causes the superconductor to enter the anomalous-metallic regime, and with increasing drive power, the resistance saturation dramatically increases.
By comparison, there is relatively little change in the resistance saturation in the quantum-critical and insulating regimes.
Likewise, the radiometry-inferred temperature of the superconductor/anomalous metal is substantially higher than in the quantum-critical or insulating regimes (Fig.~\ref{fig:addednoise}(b)) and increases more with applied power than quantum-critical or insulating regimes. The radiometry-inferred temperatures are compatible with the observed resistance saturation in all regimes.
For example, at the highest drive powers the temperature of the anomalous metal is $300-400~\mathrm{mK}$, consistent with the cryostat temperature at the onset of resistance saturation.
% Consistent with decreasing amounts of resistance saturation on the insulating side of the SIT, the sample temperatures are also lower.
% However, for each gate voltage the sample temperature increases with increasing microwave drive power.
We have also checked that a low-frequency heating tone (at $21.55~\mathrm{Hz}$) applied to the sample results in a similar increase in measured noise and correspondingly increased sample temperature.

The data in Fig.~\ref{fig:addednoise} reveal that the weakly superconducting regime is more susceptible to heating than the weakly insulating regime, suggesting a natural explanation as to why anomalous metallic resistance saturation is often observed on the superconducting side of the transition.
In a different measurement configuration, we have also studied a sample which showed a higher-temperature onset of anomalous metallic behavior, and found that it emits more radiation (see \SM~Figs.~\ref{fig:sample0}\,-\,\ref{fig:sample0_highfreq}), again consistent with a non-equilibrium origin of resistance saturation.
Although a narrow frequency band was used for noise measurements, the striking agreement between \refeq{eq:beamsplitter} and the experimental data in Figs.~\ref{fig:fig2}\,-\,\ref{fig:fig3} strongly indicates that the observed radiation power $P$ does not originate from a source locked to an array resonance frequency~\cite{barbara_stimulated_1999, cawthorne_synchronized_1999}, but rather corresponds to broadband emission.
Further supporting this viewpoint are measurements of a different sample both in the same frequency band as the main results reported here, as well as in a frequency band of $200~\mathrm{MHz}$ centered at $5~\mathrm{GHz}$ (see \SM~\refsec{sec:additional_sample}).

\section{Circuit model}

Although the temperature inference using Eq.~\ref{eq:beamsplitter} does not require a microscopic model for $S_{11}$, it is nevertheless interesting to understand the physical origin of the scattering parameters of our sample.
% It is interesting to understand the broad range of gate voltages over which reasonable impedance matching, signaled by a reduction in $S_{11}$, is observed.
% Indeed, the sample is approximately impedance matched even when its DC resistance is close to $60~\mathrm{k \Omega}$, greatly exceeding the $50~\mathrm{\Omega}$ transmission-line impedance.
To understand the basic features, we have compared our reflection data with a lossy transmission line model, see \reffig[a]{fig:figS_fits_and_inductance}.
In the model, there are infinitesimal resistors ($dR$) and inductors ($dX_\mathrm{L}$), physically representing the effect of Josephson junctions, in series with parallel capacitance ($dY_\mathrm{g}$) to ground.
A series impedance associated with the bonding pads and wires ($Z_\mathrm{in}$) is included, however we have found that it is sufficient to assume a fully real input impedance $Z_\mathrm{in} = R_\mathrm{in}$, dominated by the resistance of the aluminum bond pads in the normal state.
The details of the model and fitting procedure are included in \SM~\refsec{sec:inductance_estimates}.
The device is well described by our lossy transmission line model, with only four free fitting parameters in the normal state ($R_\mathrm{in}$, $Y_\mathrm{g}$, and the real and imaginary parts of a complex scaling factor to account for magnitude and phase offsets in our measurement setup, \reffig[b-c]{fig:figS_fits_and_inductance}).

Fixing parameters from the normal state, we can then infer a device inductance $X_\mathrm{L}$ at each gate voltage in the superconducting state  (only the device inductance $X_\mathrm{L}$, \reffig[d-e]{fig:figS_fits_and_inductance}).
The fit device inductance, shown in \reffig[f]{fig:figS_fits_and_inductance}, grows rapidly as the device is tuned from superconducting to insulating regimes, correlating strongly with the measured resistance, shown in \reffig[g]{fig:figS_fits_and_inductance}.
Above the superconducting transition temperature the device inductance is negligible, whereas at low temperatures the device response is consistent with an extremely large inductance, exceeding $10~\mathrm{\mu H}$ near the SIT and approaching $1~\mathrm{mH}$ in the insulating regime.
We interpret the extremely large inductance in the insulating phase as reflecting a diverging low-temperature inductance, $L \rightarrow \infty$ as $T \rightarrow 0$, as would physically be expected for a vanishing superfluid stiffness.

\begin{figure*}
	\centering
	\includegraphics[width=4.50in]{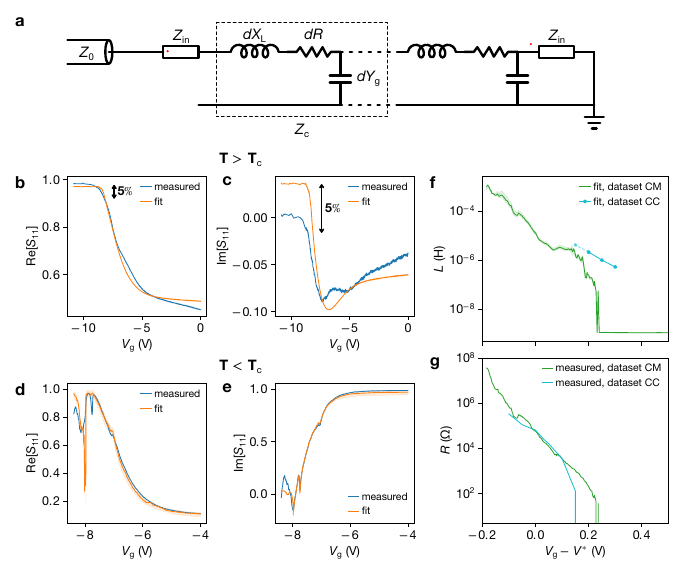}
	\caption{\textbf{a}, Circuit schematic of lossy transmission-line model describing our device.
  \textbf{b}, \textbf{c} Real and imaginary parts of $S_{11}$ above the superconducting transition temperature, both measured data and results of fitting to our model.
  At the most negative gate voltages a disagreement of 5 percent is indicated on both plots.
  \textbf{d}, \textbf{e} Real and imaginary parts of $S_{11}$ at base temperature of the cryostat, both measured data and results of fitting to our model.
  \textbf{f}, Fit inductance using our circuit model (dataset CM) and from extracted critical currents (dataset CC).
  \textbf{g} Measured zero-bias resistance corresponding to the datasets used to fit inductance.
  Dataset CM refers to the data used in the circuit model, while dataset CC refers to the data used to extract critical currents.
  In \textbf{f} and \textbf{g} the data has been shifted along the x-axis by uniformly subtracting an amount $V^*$ corresponding to the location of the SIT in transport measurements for each dataset.
  Bands on fits in \textbf{d}, \textbf{e}, \textbf{f} represent propagated uncertainties, determined by varying normal-state parameters found in \textbf{a, b} over a 5\% range, refitting at each parameter value, and taking the standard deviation of the results.}
	\label{fig:figS_fits_and_inductance}
\end{figure*} 

% Another noteworthy feature of the data in \reffig[c]{fig:fig2} is the remarkably wide range of gate voltages over which reasonable impedance matching, and therefore the possibility to measure noise, is achieved.
% We observe a rejection loss of several~dB over the gate voltage range from $-8.3~\mathrm{V}$ to $-8.75~\mathrm{V}$, corresponding to DC resistances spanning from $0$ to more than $10~\mathrm{M \Omega}$.
% In contrast to typical highly resistive samples like single electron transistors~\cite{schoelkopf_rfset_1998} where impedance matching is achieved with an external circuit, our device self-matches due to the Josephson inductance inherent to the array.
% As the gate voltage in our device is tuned towards more negative values, the coupling between adjacent superconducting islands decreases and the Josephson inductance increases, allowing the device to remain sufficiently matched throughout the SIT.
% By employing a transmission line model, we are able to reproduce the observed dependence of $S_{11}$ on $V_\mathrm{g}$ and estimate the inductance associated with the junctions as they undergo the SIT (see \SM~\refsec{sec:inductance_estimates}, particularly \reffig{fig:inductances}).
% We observe an extremely large ($>10~\mathrm{\mu H}$) and smoothly increasing inductance as the device is tuned from the superconducting to the insulating side of the apparent SIT, quantitatively demonstrating behavior similar to that seen in other systems~\cite{charpentier_first_2025}.

Summarizing up to this point, a simple model of elevated sample temperature explains the excess noise observed at the SIT, the correlation between $P$ and $|S_{11}|^2$, and the origin of anomalous metallic behavior in our device.
We emphasize that the inferred sample temperature $T_\mathrm{s}$ characterizes the measured voltage fluctuations and represents an effective temperature as defined by the fluctuation-dissipation theorem, without implying any specific microscopic interpretation.
Further, a simple circuit model can be used to understand the behavior of $|S_{11}|^2$ and make an inference of device inductance.
% Although a narrow frequency band was used for noise measurements, the striking agreement between \refeq{eq:beamsplitter} and the experimental data in Figs.~\ref{fig:fig2}\,-\,\ref{fig:fig3} strongly indicates that the observed radiation power $P$ does not originate from a source locked to an array resonance frequency~\cite{barbara_stimulated_1999, cawthorne_synchronized_1999}, but rather corresponds to broadband emission.
% Further supporting this viewpoint are measurements of a different sample both in the same frequency band as the main results reported here, as well as in a frequency band of $200~\mathrm{MHz}$ centered at $5~\mathrm{GHz}$ (see \SM~\refsec{sec:additional_sample}).

\begin{figure}
	\centering
	\includegraphics[width=3.54in]{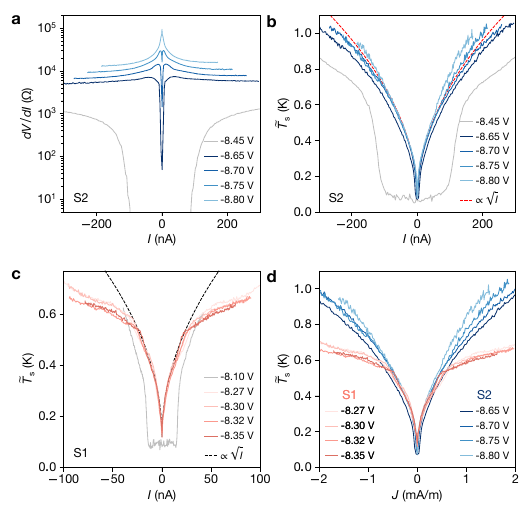}
	\caption{\textbf{a},~Differential resistance $dV/dI$ as a function of DC current, $I$, in the superconducting regime (grey) and near the superconductor-insulator transition (blue shades)
  % for different gate voltages (indicated by colors). 
	% Signatures of superconducting through insulating behavior are shown.
	\textbf{b},~Inferred sample temperature $\widetilde{T}_\mathrm{s}$ as a function of measured DC current for the same gate voltages as \textbf{a}.
	Curves near criticality collapse onto one another.
	$\sqrt{I}$ scaling is shown, motivated by predictions~\cite{dalidovich_nonlinear_2004,green_nonlinear_2005,green_current_2006,sonner_hawking_2012}.
	\textbf{c},~Inferred $\widetilde{T}_\mathrm{s}$ as a function of measured current in device S1. 
	Similar to Sample 2, curves near criticality collapse onto one another.
	$\sqrt{I}$ scaling is shown, motivated by predictions~\cite{green_current_2006,sonner_hawking_2012}.
	\textbf{d},~Inferred $\widetilde{T}_\mathrm{s}$ as a function of linear current density.
  The two different samples show a remarkable collapse for low current densities.
	Panels \textbf{a\,,\,b} show data from device S2. 
  Panel~\textbf{c} shows data from device S1.
  Panel~\textbf{d} shows data from device S1 in shades of pink and data from device S2 in shades of blue.}
	\label{fig:fig4}
\end{figure} 

\section{Finite-bias noise scaling }
We now turn to studies of noise in a non-equilibrium steady state generated by an applied bias.
In ordinary metals, this is the experimental setting for shot noise, which can reveal signatures of charge carriers and their interactions~\cite{blanter_shot_2001}.
At the superconductor-insulator critical point, it has been put forward that current fluctuations about the non-equilibrium steady states are universally determined by the dynamical critical exponent~\cite{dalidovich_nonlinear_2004,green_nonlinear_2005,green_current_2006,sonner_hawking_2012}.
To explore this intriguing concept, we have studied the behavior of our device as a function of applied bias in the quantum-critical regime.
On the superconducting side of the SIT, resistance versus current $I$ exhibits a pronounced zero-bias dip, whereas deep on the insulating side it exhibits a pronounced zero-bias peak (\reffig[a]{fig:fig4}).
Over the same gate voltage and bias range, the non-equilibrium sample temperature $\widetilde{T}_\mathrm{s}$ rises sharply with $I$.
Here we have introduced $\widetilde{T}_\mathrm{s}$, which has the same physical meaning as $T_\mathrm{s}$ (sample temperature defined by Eq.~\ref{eq:beamsplitter}), but uses a different method for fixing $T_\mathrm{cryo}$ (see \SM~\refsec{sec:cal_sn}).
% as an alternative inference of sample temperature in order to cancel out weak, time-dependent added noise drifts of the measurement chain (see \SM~\refsec{sec:calibration}, particularly Figs.~\ref{fig:responsivity}\,-\,\ref{fig:fig_calibration}).

In contrast to resistance, whose zero-bias value varies by three orders of magnitude over the measured range of gate voltages, $\widetilde{T}_\mathrm{s}$ is nearly gate voltage independent, and approximately collapses onto a single nonlinear curve (\reffig[b]{fig:fig4}), compatible with the theoretically predicted $\sqrt{I}$ scaling \cite{dalidovich_nonlinear_2004,green_nonlinear_2005,green_current_2006,sonner_hawking_2012}.
The collapse only occurs when plotting $\widetilde{T}_\mathrm{s}$ versus $I$; other noise metrics such as the output noise power spectral density $P$ (\reffig[c]{fig:fig4}) or dependent variables such as voltage $V$, or Joule power $I V$ do not exhibit a collapse (\reffig{fig:nocollapse}).
We emphasize that the scaling behavior of the samples' temperature $\widetilde{T}_\mathrm{s}$ irrespective of impedance suggests that the non-equilibrium behavior is not affected by device impedance, in contrast to resonance effects identified in earlier work with two-dimensional arrays \cite{barbara_stimulated_1999}.
A qualitatively similar collapse is observed in a sample with a different geometry (\reffig[c]{fig:fig4}), confirming that the scaling phenomena is reproducible across samples.
In fact, the data across both samples collapse at low bias when plotted as a function of the current density (\reffig[d]{fig:fig4}), hinting at the existence of a universal scaling non-equilibrium behavior at the superconductor-insulator transition.

The observed scaling behavior of $\widetilde{T}_\mathrm{s}$ is reminiscent of theoretical predictions of universal non-equilibrium behavior near the SIT~\cite{dalidovich_nonlinear_2004,green_nonlinear_2005,green_current_2006, sonner_hawking_2012}.
These theories predict that current fluctuations are described by an effective temperature, mimicking the fluctuation-dissipation relation even though the system is profoundly non-equilibrium.
Furthermore, the effective temperature is predicted to exhibit a universal scaling square-root scaling with applied bias, similar to the experimental data reported in \reffig{fig:fig4}.
Two additional experimental observations support a connection between non-equilibrium steady states and equivalent temperature.
First, scaling in $\widetilde{T}_\mathrm{s}$ in \reffig[b,\,c]{fig:fig4} occurs over a comparable temperature range to temperature-independence of near-critical curves in \reffig[b]{fig:fig1}.
Second, zero-bias and high-bias differential resistance are comparable near the superconductor-insulator transition, where differential resistance is temperature independent (see \SM~\refsec{sec:finite_bias}).
However, theories predict that the effective temperature scales as a function of applied voltage, in contrast to the experimentally observed scaling with current.
It is unclear if this discrepancy is crucial, as these theories also predict a universal high-bias conductance linking current and voltage~\cite{dalidovich_nonlinear_2004,green_nonlinear_2005,green_nonlinear_2005}, while we observe a strong bias-dependent resistance.
A new and interesting perspective can also be given by recent theoretical work, which shows that broadband radiation can be expected even for  a single Josephson junction in a high-impedance environment \cite{kurilovich_quantum_2025}.
A full understanding of non-equilibrium responses near the SIT will require both further experimental work, for example varying sample sizes and model systems, as well as further theoretical input.
However, these initial results indicate the appealing possibility of universal, non-equilibrium behavior near quantum criticality.

\section{Summary}

Before concluding, several points need further clarification.
Of relevance to \reffig[b]{fig:fig2}, we emphasize that the observed gate-voltage dependence of $P$ is primarily explained by impedance matching effects (in conjunction with elevated sample temperature); for extremal gate voltages where the device is mismatched, we cannot make an experimental statement about the sample temperature.
Of relevance to \reffig{fig:fig4}, we give some additional discussion of the heat-sink mechanism in our samples.
One possibility, supported by recent studies of single junctions~\cite{karimi_bolometric_2024}, is that the primary thermalization mechanism is radiative.
The observed scaling with bias current $I$ might then suggest a picture of internal thermalization with Josephson radiation, which eventually escapes through the leads.
When the device is driven normal by a magnetic field, our data are consistent with conventional thermalization limited by the electron-phonon bottleneck (see \SM~\refsec{sec:magnetic_field}).
Expressing total noise in units of current spectral density $S_I$, we estimate Fano factors $F$ defined by $S_I = 2eFI$~\cite{blanter_shot_2001} in the range $5-10$, suggesting that heat does not leak and dissipate underneath the superconducting contacts~\cite{denisov_heat-mode_2022}.
We note that similar Fano factors were observed in earlier measurements of SNS junctions~\cite{dieleman_observation_1997,ronen_charge_2016,hoss_multiple_2000}, which could microscopically arise from a combination of multiple Andreev reflections, Josephson radiation, or from the overheating of normal, diffusive regions between superconducting islands~\cite{courtois_origin_2008}.
Also of relevance to \reffig{fig:fig4}, we note that in contrast to the low-frequency resistance, the high-frequency conductivity is only weakly gate dependent.
Weak gate dependence makes it experimentally convenient to extract the equivalent temperature from high-frequency noise data.
Finally, we note that weak, stray magnetic fields could plausibly create dilute vortices, which tend to give dissipation and heating at low temperature.
Additional studies with high levels of shielding or field compensation would be interesting.

Summarizing, we have implemented microwave radiometry as a powerful probe of behavior near the superconductor-insulator transition.
We find that anomalous metallic behavior in our system correlates with a failure to thermalize with the cryostat.
We have also studied heating as a function of applied bias, observing scaling behavior that was anticipated based on Keldysh, Boltzmann, and gravity-gauge duality approaches~\cite{dalidovich_nonlinear_2004,green_nonlinear_2005,green_current_2006,sonner_hawking_2012}.

It would be interesting to perform further thermometry measurements in a wider variety of model systems.
It is possible that anomalous metallic saturation is associated with non-equilibrium behavior in some cases, but that in other cases there is equilibrium metallic behavior as $T\rightarrow 0$~\cite{feigelman_quantum_1998,spivak_quantum_2001,spivak_theory_2008,kapitulnik_colloquium_2019}.
Recent work in indium oxide \cite{haug_excess_2024} also found deviations from equilibrium behavior near the superconductor-insulator transition, raising the possibility that lack of equilibration is generic across systems.
The experimental approach demonstrated here, combined with recently-demonstrated on-chip bolometry~\cite{karimi_bolometric_2024}, sets the stage for a new exploration of non-equilibrium behavior in Josephson arrays, weak superconductors, and other near-critical systems.

\textbf{Acknowledgements:} We gratefully acknowledge feedback on the pre-print from Charles Marcus, Vadim Khrapai, Joel Moore, Andrew Green, Shivaji Sondhi, and Rufus Boyack.
This work was primarily supported by the NOMIS foundation.
This work was partially supported by the University of Chicago Materials Research Science and Engineering Center, which is funded by the National Science Foundation under award number DMR-2011854, and by the SFB Q-M\&S funded by the Austrian Science Fund (FWF).
We acknowledge technical support from the Nanofabrication Facility and the MIBA machine shop at IST Austria.

\textbf{Data Availability:} The data that support the findings of this article are openly available~\cite{zenodo_dataset_2026}.

\nocite{*}

\bibliography{main}% Produces the bibliography via BibTeX.

@misc{ zenodo_dataset_2026,
  title        = {Data for "Microwave radiometry of a quantum-critical, hybrid Josephson array"},
  year         = {2026},
  howpublished = {\url{https://doi.org/10.5281/zenodo.18511839}},
  note         = {{Zenodo}},
  doi          = {10.5281/zenodo.18511839},
}

@article{ abrahams_scaling_1979,
  title = {Scaling Theory of Localization: Absence of Quantum Diffusion in Two Dimensions},
  author = {Abrahams, E. and Anderson, P. W. and Licciardello, D. C. and Ramakrishnan, T. V.},
  journal = {Phys. Rev. Lett.},
  volume = {42},
  issue = {10},
  pages = {673--676},
  numpages = {0},
  year = {1979},
  month = {Mar},
  publisher = {American Physical Society},
  doi = {10.1103/PhysRevLett.42.673},
  url = {https://link.aps.org/doi/10.1103/PhysRevLett.42.673}
}

@article{ sacepe_quantum_2020,
	author = {Sac{\'e}p{\'e}, Benjamin and Feigel'man, Mikhail and Klapwijk, Teunis M.},
	date = {2020/07/01},
	doi = {10.1038/s41567-020-0905-x},
	id = {Sac{\'e}p{\'e}2020},
	isbn = {1745-2481},
	journal = {Nature Physics},
	number = {7},
	pages = {734--746},
	title = {Quantum breakdown of superconductivity in low-dimensional materials},
	url = {https://doi.org/10.1038/s41567-020-0905-x},
	volume = {16},
	year = {2020}}

@article{ majumdar_universality_2025,
	author = {Majumdar, Aniket and Chadha, Nisarg and Pal, Pritam and Gugnani, Akash and Ghawri, Bhaskar and Watanabe, Kenji and Taniguchi, Takashi and Mukerjee, Subroto and Ghosh, Arindam},
	date = {2025/08/13},
	doi = {10.1038/s41567-025-02972-z},
	id = {Majumdar2025},
	isbn = {1745-2481},
	journal = {Nature Physics},
	title = {Universality in quantum critical flow of charge and heat in ultraclean graphene},
	url = {https://doi.org/10.1038/s41567-025-02972-z},
	year = {2025}}

@article{ chen_shot_2023,
author = {Liyang Chen  and Dale T. Lowder  and Emine Bakali  and Aaron Maxwell Andrews  and Werner Schrenk  and Monika Waas  and Robert Svagera  and Gaku Eguchi  and Lukas Prochaska  and Yiming Wang  and Chandan Setty  and Shouvik Sur  and Qimiao Si  and Silke Paschen  and Douglas Natelson },
title = {Shot noise in a strange metal},
journal = {Science},
volume = {382},
number = {6673},
pages = {907-911},
year = {2023},
doi = {10.1126/science.abq6100},
URL = {https://www.science.org/doi/abs/10.1126/science.abq6100},
eprint = {https://www.science.org/doi/pdf/10.1126/science.abq6100}}

@article{ kurilovich_quantum_2025,
	author = {Kurilovich, Vladislav D. and Remez, Benjamin and Glazman, Leonid I.},
	date = {2025/02/05},
	doi = {10.1038/s41467-025-56411-x},
	id = {Kurilovich2025},
	isbn = {2041-1723},
	journal = {Nature Communications},
	number = {1},
	pages = {1384},
	title = {Quantum theory of Bloch oscillations in a resistively shunted transmon},
	url = {https://doi.org/10.1038/s41467-025-56411-x},
	volume = {16},
	year = {2025}}

@article{ chatterjee_semiconductor_2021,
	author = {Chatterjee, Anasua and Stevenson, Paul and De Franceschi, Silvano and Morello, Andrea and de Leon, Nathalie P. and Kuemmeth, Ferdinand},
	date = {2021/03/01},
	doi = {10.1038/s42254-021-00283-9},
	id = {Chatterjee2021},
	isbn = {2522-5820},
	journal = {Nature Reviews Physics},
	number = {3},
	pages = {157--177},
	title = {Semiconductor qubits in practice},
	url = {https://doi.org/10.1038/s42254-021-00283-9},
	volume = {3},
	year = {2021}}

@article{ sasmal_voltage-tuned_2025,
  title = {Voltage-tuned anomalous-metal to metal transition in hybrid Josephson junction arrays},
  author = {S. Sasmal and M. Efthymiou-Tsironi and G. Nagda and E. Fugl and L. L. Olsen and F. Krizek and C. M. Marcus and S. Vaitiekėnas},
  year = {2025},
  journal = {arXiv:2505.12536},
  doi = {10.48550/arXiv.2505.12536}
}

@article{ haug_excess_2024,
  title = {Excess noise in the anomalous metallic phase in amorphous indium oxide},
  author = {Haug, Andr\'e and Shahar, Dan},
  journal = {Phys. Rev. B},
  volume = {109},
  issue = {1},
  pages = {014514},
  numpages = {5},
  year = {2024},
  month = {Jan},
  publisher = {American Physical Society},
  doi = {10.1103/PhysRevB.109.014514},
  url = {https://link.aps.org/doi/10.1103/PhysRevB.109.014514}
}

@article{ giazotto_opportunities_2006,
  title = {Opportunities for mesoscopics in thermometry and refrigeration: Physics and applications},
  author = {Giazotto, Francesco and Heikkil\"a, Tero T. and Luukanen, Arttu and Savin, Alexander M. and Pekola, Jukka P.},
  journal = {Rev. Mod. Phys.},
  volume = {78},
  issue = {1},
  pages = {217--274},
  numpages = {0},
  year = {2006},
  month = {Mar},
  publisher = {American Physical Society},
  doi = {10.1103/RevModPhys.78.217},
  url = {https://link.aps.org/doi/10.1103/RevModPhys.78.217}
}

@article{ boettcher_Berezinskii_2022,
      title = {The Berezinskii-Kosterlitz-Thouless Transition and Anomalous Metallic Phase in a Hybrid Josephson Junction Array}, 
      author = {C. G. L. Bøttcher and F. Nichele and J. Shabani and C. J. Palmstrøm and C. M. Marcus},
      year = {2022},
      eprint = {2210.00318},
      archivePrefix = {arXiv},
      journal = {arXiv preprint},
      primaryClass = {cond-mat.mes-hall},
      url = {https://arxiv.org/abs/2210.00318}
}

@article{ haviland_onset_1989,
  title = {Onset of superconductivity in the two-dimensional limit},
  author = {Haviland, D. B. and Liu, Y. and Goldman, A. M.},
  journal = {Phys. Rev. Lett.},
  volume = {62},
  issue = {18},
  pages = {2180--2183},
  numpages = {0},
  year = {1989},
  month = {May},
  publisher = {American Physical Society},
  doi = {10.1103/PhysRevLett.62.2180},
  url = {https://link.aps.org/doi/10.1103/PhysRevLett.62.2180}
}

@article{ spivak_theory_2008,
  title = {Theory of quantum metal to superconductor transitions in highly conducting systems},
  author = {Spivak, B. and Oreto, P. and Kivelson, S. A.},
  journal = {Phys. Rev. B},
  volume = {77},
  issue = {21},
  pages = {214523},
  numpages = {18},
  year = {2008},
  month = {Jun},
  publisher = {American Physical Society},
  doi = {10.1103/PhysRevB.77.214523},
  url = {https://link.aps.org/doi/10.1103/PhysRevB.77.214523}
}

@article{ spivak_quantum_2001,
  title = {Quantum superconductor-metal transition},
  author = {Spivak, B. and Zyuzin, A. and Hruska, M.},
  journal = {Phys. Rev. B},
  volume = {64},
  issue = {13},
  pages = {132502},
  numpages = {4},
  year = {2001},
  month = {Aug},
  publisher = {American Physical Society},
  doi = {10.1103/PhysRevB.64.132502},
  url = {https://link.aps.org/doi/10.1103/PhysRevB.64.132502}
}

@article{ feigelman_quantum_1998,
title = {Quantum superconductor–metal transition in a 2D proximity-coupled array},
journal = {Chemical Physics},
volume = {235},
number = {1},
pages = {107-114},
year = {1998},
issn = {0301-0104},
doi = {https://doi.org/10.1016/S0301-0104(98)00075-5},
url = {https://www.sciencedirect.com/science/article/pii/S0301010498000755},
author = {M.V. Feigel'man and A.I. Larkin}
}

@article{mayer_superconducting_2019,
    author = {Mayer, William and Yuan, Joseph and Wickramasinghe, Kaushini S. and Nguyen, Tri and Dartiailh, Matthieu C. and Shabani, Javad},
    title = "{Superconducting proximity effect in epitaxial Al-InAs heterostructures}",
    journal = {Applied Physics Letters},
    volume = {114},
    number = {10},
    pages = {103104},
    year = {2019},
    month = {03},
    issn = {0003-6951},
    doi = {10.1063/1.5067363},
    url = {https://doi.org/10.1063/1.5067363},
    eprint = {https://pubs.aip.org/aip/apl/article-pdf/doi/10.1063/1.5067363/19772789/103104\_1\_online.pdf},
}

@article{ castellanos-beltran_amplification_2008,
	author = {Castellanos-Beltran, M. A. and Irwin, K. D. and Hilton, G. C. and Vale, L. R. and Lehnert, K. W.},
	date = {2008/12/01},
	doi = {10.1038/nphys1090},
	id = {Castellanos-Beltran2008},
	isbn = {1745-2481},
	journal = {Nature Physics},
	number = {12},
	pages = {929--931},
	title = {Amplification and squeezing of quantum noise with a tunable Josephson metamaterial},
	url = {https://doi.org/10.1038/nphys1090},
	volume = {4},
	year = {2008}}

@article{ dicke_atmospheric_1946,
  title = {Atmospheric Absorption Measurements with a Microwave Radiometer},
  author = {Dicke, Robert H. and Beringer, Robert and Kyhl, Robert L. and Vane, A. B.},
  journal = {Phys. Rev.},
  volume = {70},
  issue = {5-6},
  pages = {340--348},
  numpages = {0},
  year = {1946},
  month = {Sep},
  publisher = {American Physical Society},
  doi = {10.1103/PhysRev.70.340},
  url = {https://link.aps.org/doi/10.1103/PhysRev.70.340}
}

@article{ brubaker_first_2017,
  title = {First Results from a Microwave Cavity Axion Search at $24\text{ }\text{ }\ensuremath{\mu}\mathrm{eV}$},
  author = {Brubaker, B. M. and Zhong, L. and Gurevich, Y. V. and Cahn, S. B. and Lamoreaux, S. K. and Simanovskaia, M. and Root, J. R. and Lewis, S. M. and Al Kenany, S. and Backes, K. M. and Urdinaran, I. and Rapidis, N. M. and Shokair, T. M. and van Bibber, K. A. and Palken, D. A. and Malnou, M. and Kindel, W. F. and Anil, M. A. and Lehnert, K. W. and Carosi, G.},
  journal = {Phys. Rev. Lett.},
  volume = {118},
  issue = {6},
  pages = {061302},
  numpages = {5},
  year = {2017},
  month = {Feb},
  publisher = {American Physical Society},
  doi = {10.1103/PhysRevLett.118.061302},
  url = {https://link.aps.org/doi/10.1103/PhysRevLett.118.061302}
}

@article{ denisov_heat-mode_2022,
AUTHOR = {Denisov, Artem and Bubis, Anton and Piatrusha, Stanislau and Titova, Nadezhda and Nasibulin, Albert and Becker, Jonathan and Treu, Julian and Ruhstorfer, Daniel and Koblmüller, Gregor and Tikhonov, Evgeny and Khrapai, Vadim},
TITLE = {Heat-Mode Excitation in a Proximity Superconductor},
JOURNAL = {Nanomaterials},
VOLUME = {12},
YEAR = {2022},
NUMBER = {9},
ARTICLE-NUMBER = {1461},
URL = {https://www.mdpi.com/2079-4991/12/9/1461},
PubMedID = {35564170},
ISSN = {2079-4991},
DOI = {10.3390/nano12091461}
}

@article{ steinbach_observation_1996,
  title = {Observation of Hot-Electron Shot Noise in a Metallic Resistor},
  author = {Steinbach, Andrew H. and Martinis, John M. and Devoret, Michel H.},
  journal = {Phys. Rev. Lett.},
  volume = {76},
  issue = {20},
  pages = {3806--3809},
  numpages = {0},
  year = {1996},
  month = {May},
  publisher = {American Physical Society},
  doi = {10.1103/PhysRevLett.76.3806},
  url = {https://link.aps.org/doi/10.1103/PhysRevLett.76.3806}
}

@article{ green_current_2006,
  title = {Current Noise in the Vicinity of the 2D Superconductor-Insulator Quantum Critical Point},
  author = {Green, A. G. and Moore, J. E. and Sondhi, S. L. and Vishwanath, A.},
  journal = {Phys. Rev. Lett.},
  volume = {97},
  issue = {22},
  pages = {227003},
  numpages = {4},
  year = {2006},
  month = {Dec},
  publisher = {American Physical Society},
  doi = {10.1103/PhysRevLett.97.227003},
  url = {https://link.aps.org/doi/10.1103/PhysRevLett.97.227003}
}

@article{ sonner_hawking_2012,
  title = {Hawking Radiation and Nonequilibrium Quantum Critical Current Noise},
  author = {Sonner, Julian and Green, A. G.},
  journal = {Phys. Rev. Lett.},
  volume = {109},
  issue = {9},
  pages = {091601},
  numpages = {5},
  year = {2012},
  month = {Aug},
  publisher = {American Physical Society},
  doi = {10.1103/PhysRevLett.109.091601},
  url = {https://link.aps.org/doi/10.1103/PhysRevLett.109.091601}
}

@article{ jaeger_onset_1989,
  title = {Onset of superconductivity in ultrathin granular metal films},
  author = {Jaeger, H. M. and Haviland, D. B. and Orr, B. G. and Goldman, A. M.},
  journal = {Phys. Rev. B},
  volume = {40},
  issue = {1},
  pages = {182--196},
  numpages = {0},
  year = {1989},
  month = {Jul},
  publisher = {American Physical Society},
  doi = {10.1103/PhysRevB.40.182},
  url = {https://link.aps.org/doi/10.1103/PhysRevB.40.182}
}

@article{ ephron_observation_1996,
  title = {Observation of Quantum Dissipation in the Vortex State of a Highly Disordered Superconducting Thin Film},
  author = {Ephron, D. and Yazdani, A. and Kapitulnik, A. and Beasley, M. R.},
  journal = {Phys. Rev. Lett.},
  volume = {76},
  issue = {9},
  pages = {1529--1532},
  numpages = {0},
  year = {1996},
  month = {Feb},
  publisher = {American Physical Society},
  doi = {10.1103/PhysRevLett.76.1529},
  url = {https://link.aps.org/doi/10.1103/PhysRevLett.76.1529}
}

@article{ rimberg_dissipation_1997,
  title = {Dissipation-Driven Superconductor-Insulator Transition in a Two-Dimensional Josephson-Junction Array},
  author = {Rimberg, A. J. and Ho, T. R. and Kurdak, \ifmmode \mbox{\c{C}}\else \c{C}\fi{}. and Clarke, John and Campman, K. L. and Gossard, A. C.},
  journal = {Phys. Rev. Lett.},
  volume = {78},
  issue = {13},
  pages = {2632--2635},
  numpages = {0},
  year = {1997},
  month = {Mar},
  publisher = {American Physical Society},
  doi = {10.1103/PhysRevLett.78.2632},
  url = {https://link.aps.org/doi/10.1103/PhysRevLett.78.2632}
}

@Article{ boettcher_superconducting_2018,
    author = {B{\o}ttcher, C. G. L. and Nichele, F. and Kjaergaard, M. and Suominen, H. J. and Shabani, J. and Palmstr{\o}m, C. J. and Marcus, C. M.},
    title = {Superconducting, insulating and anomalous metallic regimes in a gated two-dimensional semiconductor--superconductor array},
    journal = {Nature Physics},
    year = {2018},
    month = {Nov},
    volume = {14},
    number = {11},
    pages = {1138-1144},
    issn = {1745-2481},
    doi = {10.1038/s41567-018-0259-9},
    url = {https://doi.org/10.1038/s41567-018-0259-9}
}

@article{ kapitulnik_colloquium_2019,
  title = {Colloquium: Anomalous metals: Failed superconductors},
  author = {Kapitulnik, Aharon and Kivelson, Steven A. and Spivak, Boris},
  journal = {Rev. Mod. Phys.},
  volume = {91},
  issue = {1},
  pages = {011002},
  numpages = {25},
  year = {2019},
  month = {Jan},
  publisher = {American Physical Society},
  doi = {10.1103/RevModPhys.91.011002},
  url = {https://link.aps.org/doi/10.1103/RevModPhys.91.011002}
}

@article{ tamir_sensitivity_2019,
    author = {I. Tamir  and A. Benyamini  and E. J. Telford  and F. Gorniaczyk  and A. Doron  and T. Levinson  and D. Wang  and F. Gay  and B. Sacépé  and J. Hone  and K. Watanabe  and T. Taniguchi  and C. R. Dean  and A. N. Pasupathy  and D. Shahar },
    title = {Sensitivity of the superconducting state in thin films},
    journal = {Science Advances},
    volume = {5},
    number = {3},
    pages = {eaau3826},
    year = {2019},
    doi = {10.1126/sciadv.aau3826},
    URL = {https://www.science.org/doi/abs/10.1126/sciadv.aau3826},
    eprint = {https://www.science.org/doi/pdf/10.1126/sciadv.aau3826},
}

@article{ goldman_superconductor_2003,
    title = {Superconductor–insulator transitions in the two-dimensional limit},
    journal = {Physica E: Low-dimensional Systems and Nanostructures},
    volume = {18},
    number = {1},
    pages = {1-6},
    year = {2003},
    note = {23rd International Conference on Low Temperature Physics (LT23)},
    issn = {1386-9477},
    doi = {https://doi.org/10.1016/S1386-9477(02)00932-3},
    url = {https://www.sciencedirect.com/science/article/pii/S1386947702009323},
    author = {A.M. Goldman},
}

@Article{ ienaga_broadened_2024,
    author={Ienaga, Koichiro and Tamoto, Yutaka and Yoda, Masahiro and Yoshimura, Yuki and Ishigami, Takahiro and Okuma, Satoshi},
    title={Broadened quantum critical ground state in a disordered superconducting thin film},
    journal={Nature Communications},
    year={2024},
    month={Mar},
    day={16},
    volume={15},
    number={1},
    pages={2388},
    issn={2041-1723},
    doi={10.1038/s41467-024-46628-7},
    url={https://doi.org/10.1038/s41467-024-46628-7}
}

@article{ breznay_self_2015,
    author = {Nicholas P. Breznay  and Myles A. Steiner  and Steven Allan Kivelson  and Aharon Kapitulnik },
    title = {Self-duality and a Hall-insulator phase near the superconductor-to-insulator transition in indium-oxide films},
    journal = {Proceedings of the National Academy of Sciences},
    volume = {113},
    number = {2},
    pages = {280-285},
    year = {2016},
    doi = {10.1073/pnas.1522435113},
    URL = {https://www.pnas.org/doi/abs/10.1073/pnas.1522435113},
    eprint = {https://www.pnas.org/doi/pdf/10.1073/pnas.1522435113}
}

@article{ han_collapse_2014,
    author={Han, Zheng and Allain, Adrien and Arjmandi-Tash, Hadi and Tikhonov, Konstantin and Feigel'man, Mikhail and Sac{\'e}p{\'e}, Benjamin and Bouchiat, Vincent},
    title={Collapse of superconductivity in a hybrid tin--graphene Josephson junction array},
    journal={Nature Physics},
    year={2014},
    month={May},
    day={01},
    volume={10},
    number={5},
    pages={380-386},
    issn={1745-2481},
    doi={10.1038/nphys2929},
    url={https://doi.org/10.1038/nphys2929}
}

@article{ boettcher_dynamical_2023,
  title = {Dynamical vortex transitions in a gate-tunable two-dimensional Josephson junction array},
  author = {B\o{}ttcher, C. G. L. and Nichele, F. and Shabani, J. and Palmstr\o{}m, C. J. and Marcus, C. M.},
  journal = {Phys. Rev. B},
  volume = {108},
  issue = {13},
  pages = {134517},
  numpages = {8},
  year = {2023},
  month = {Oct},
  publisher = {American Physical Society},
  doi = {10.1103/PhysRevB.108.134517},
  url = {https://link.aps.org/doi/10.1103/PhysRevB.108.134517}
}

@article{ shin_effect_2020,
  title = {Effect of external electromagnetic radiation on the anomalous metallic behavior in superconducting Ta thin films},
  author = {Shin, Junghyun and Park, Sungyu and Kim, Eunseong},
  journal = {Phys. Rev. B},
  volume = {102},
  issue = {18},
  pages = {184501},
  numpages = {6},
  year = {2020},
  month = {Nov},
  publisher = {American Physical Society},
  doi = {10.1103/PhysRevB.102.184501},
  url = {https://link.aps.org/doi/10.1103/PhysRevB.102.184501}
}

@article{ crauste_thickness_2009,
  doi = {10.1088/1742-6596/150/4/042019},
  url = {https://dx.doi.org/10.1088/1742-6596/150/4/042019%7},
  year = {2009},
  month = {mar},
  publisher = {},
  volume = {150},
  number = {4},
  pages = {042019},
  author = {O Crauste and  C A Marrache-Kikuchi and  L Bergé and  D Stanescu and  L Dumoulin},
  title = {Thickness dependence of the superconductivity in thin disordered NbSi films},
  journal = {Journal of Physics: Conference Series}
}

@article{ breznay_particle_2017,
  author = {Breznay, Nicholas P. and Kapitulnik, Aharon},
  title = {Particle-hole symmetry reveals failed superconductivity in the metallic phase of two-dimensional superconducting films},
  journal = {Science Advances},
  year = {2017},
  publisher = {American Association for the Advancement of Science},
  volume = {3},
  number = {9},
  pages = {e1700612},
  doi = {10.1126/sciadv.1700612},
  url = {https://doi.org/10.1126/sciadv.1700612}
}

@article{ shabani_two-dimensional_2016,
  title = {Two-dimensional epitaxial superconductor-semiconductor heterostructures: A platform for topological superconducting networks},
  author = {Shabani, J. and Kjaergaard, M. and Suominen, H. J. and Kim, Younghyun and Nichele, F. and Pakrouski, K. and Stankevic, T. and Lutchyn, R. M. and Krogstrup, P. and Feidenhans'l, R. and Kraemer, S. and Nayak, C. and Troyer, M. and Marcus, C. M. and Palmstr\o{}m, C. J.},
  journal = {Phys. Rev. B},
  volume = {93},
  issue = {15},
  pages = {155402},
  numpages = {6},
  year = {2016},
  month = {Apr},
  publisher = {American Physical Society},
  doi = {10.1103/PhysRevB.93.155402},
  url = {https://link.aps.org/doi/10.1103/PhysRevB.93.155402}
}

@article{ dieleman_observation_1997,
  title = {Observation of Andreev Reflection Enhanced Shot Noise},
  author = {Dieleman, P. and Bukkems, H. G. and Klapwijk, T. M. and Schicke, M. and Gundlach, K. H.},
  journal = {Phys. Rev. Lett.},
  volume = {79},
  issue = {18},
  pages = {3486--3489},
  numpages = {0},
  year = {1997},
  month = {Nov},
  publisher = {American Physical Society},
  doi = {10.1103/PhysRevLett.79.3486},
  url = {https://link.aps.org/doi/10.1103/PhysRevLett.79.3486}
}

@article{ ronen_charge_2016,
  author = {Yuval Ronen  and Yonatan Cohen  and Jung-Hyun Kang  and Arbel Haim  and Maria-Theresa Rieder  and Moty Heiblum  and Diana Mahalu  and Hadas Shtrikman },
  title = {Charge of a quasiparticle in a superconductor},
  journal = {Proceedings of the National Academy of Sciences},
  volume = {113},
  number = {7},
  pages = {1743-1748},
  year = {2016},
  doi = {10.1073/pnas.1515173113},
  URL = {https://www.pnas.org/doi/abs/10.1073/pnas.1515173113},
  eprint = {https://www.pnas.org/doi/pdf/10.1073/pnas.1515173113},
}

@article{ blanter_shot_2001,
  title = {Shot noise in mesoscopic conductors},
  journal = {Physics Reports},
  volume = {336},
  number = {1},
  pages = {1-166},
  year = {2000},
  issn = {0370-1573},
  doi = {https://doi.org/10.1016/S0370-1573(99)00123-4},
  url = {https://www.sciencedirect.com/science/article/pii/S0370157399001234},
  author = {Ya.M. Blanter and M. Büttiker}
}

@article{ hoss_multiple_2000,
  title = {Multiple Andreev reflection and giant excess noise in diffusive superconductor/normal-metal/superconductor junctions},
  author = {Hoss, T. and Strunk, C. and Nussbaumer, T. and Huber, R. and Staufer, U. and Sch\"onenberger, C.},
  journal = {Phys. Rev. B},
  volume = {62},
  issue = {6},
  pages = {4079--4085},
  numpages = {0},
  year = {2000},
  month = {Aug},
  publisher = {American Physical Society},
  doi = {10.1103/PhysRevB.62.4079},
  url = {https://link.aps.org/doi/10.1103/PhysRevB.62.4079}
}

@article{ mithun_dynamical_2019,
  title = {Dynamical Glass and Ergodization Times in Classical Josephson Junction Chains},
  author = {Mithun, Thudiyangal and Danieli, Carlo and Kati, Yagmur and Flach, Sergej},
  journal = {Phys. Rev. Lett.},
  volume = {122},
  issue = {5},
  pages = {054102},
  numpages = {6},
  year = {2019},
  month = {Feb},
  publisher = {American Physical Society},
  doi = {10.1103/PhysRevLett.122.054102},
  url = {https://link.aps.org/doi/10.1103/PhysRevLett.122.054102}
}

@article{ mithun_weakly_2018,
  title = {Weakly Nonergodic Dynamics in the Gross-Pitaevskii Lattice},
  author = {Mithun, Thudiyangal and Kati, Yagmur and Danieli, Carlo and Flach, Sergej},
  journal = {Phys. Rev. Lett.},
  volume = {120},
  issue = {18},
  pages = {184101},
  numpages = {6},
  year = {2018},
  month = {May},
  publisher = {American Physical Society},
  doi = {10.1103/PhysRevLett.120.184101},
  url = {https://link.aps.org/doi/10.1103/PhysRevLett.120.184101}
}

@article{ pino_nonergodic_2016,
author = {Manuel Pino  and Lev B. Ioffe  and Boris L. Altshuler },
title = {Nonergodic metallic and insulating phases of Josephson junction chains},
journal = {Proceedings of the National Academy of Sciences},
volume = {113},
number = {3},
pages = {536-541},
year = {2016},
doi = {10.1073/pnas.1520033113},
URL = {https://www.pnas.org/doi/abs/10.1073/pnas.1520033113},
eprint = {https://www.pnas.org/doi/pdf/10.1073/pnas.1520033113}
}

@article{ pino_multifractal_2017,
  title = {Multifractal metal in a disordered Josephson junctions array},
  author = {Pino, M. and Kravtsov, V. E. and Altshuler, B. L. and Ioffe, L. B.},
  journal = {Phys. Rev. B},
  volume = {96},
  issue = {21},
  pages = {214205},
  numpages = {10},
  year = {2017},
  month = {Dec},
  publisher = {American Physical Society},
  doi = {10.1103/PhysRevB.96.214205},
  url = {https://link.aps.org/doi/10.1103/PhysRevB.96.214205}
}

@article{ cha_universal_1991,
  title = {Universal conductivity of two-dimensional films at the superconductor-insulator transition},
  author = {Cha, Min-Chul and Fisher, Matthew P. A. and Girvin, S. M. and Wallin, Mats and Young, A. Peter},
  journal = {Phys. Rev. B},
  volume = {44},
  issue = {13},
  pages = {6883--6902},
  numpages = {0},
  year = {1991},
  month = {Oct},
  publisher = {American Physical Society},
  doi = {10.1103/PhysRevB.44.6883},
  url = {https://link.aps.org/doi/10.1103/PhysRevB.44.6883}
}

@article{ courtois_origin_2008,
  title = {Origin of Hysteresis in a Proximity Josephson Junction},
  author = {Courtois, H. and Meschke, M. and Peltonen, J. T. and Pekola, J. P.},
  journal = {Phys. Rev. Lett.},
  volume = {101},
  issue = {6},
  pages = {067002},
  numpages = {4},
  year = {2008},
  month = {Aug},
  publisher = {American Physical Society},
  doi = {10.1103/PhysRevLett.101.067002},
  url = {https://link.aps.org/doi/10.1103/PhysRevLett.101.067002}
}

@article{ karimi_bolometric_2024,
	author = {Karimi, Bayan and Steffensen, Gorm Ole and Higginbotham, Andrew P. and Marcus, Charles M. and Levy Yeyati, Alfredo and Pekola, Jukka P.},
	date = {2024/08/22},
	doi = {10.1038/s41565-024-01770-7},
	id = {Karimi2024},
	isbn = {1748-3395},
	journal = {Nature Nanotechnology},
	title = {Bolometric detection of Josephson radiation},
	url = {https://doi.org/10.1038/s41565-024-01770-7},
	year = {2024}}

@article{ swanson_dynamical_2014,
  title = {Dynamical Conductivity across the Disorder-Tuned Superconductor-Insulator Transition},
  author = {Swanson, Mason and Loh, Yen Lee and Randeria, Mohit and Trivedi, Nandini},
  journal = {Phys. Rev. X},
  volume = {4},
  issue = {2},
  pages = {021007},
  numpages = {10},
  year = {2014},
  month = {Apr},
  publisher = {American Physical Society},
  doi = {10.1103/PhysRevX.4.021007},
  url = {https://link.aps.org/doi/10.1103/PhysRevX.4.021007}
}

@article{ eley_approaching_2012,
	author = {Eley, Serena and Gopalakrishnan, Sarang and Goldbart, Paul M. and Mason, Nadya},
	date = {2012/01/01},
	doi = {10.1038/nphys2154},
	id = {Eley2012},
	isbn = {1745-2481},
	journal = {Nature Physics},
	number = {1},
	pages = {59--62},
	title = {Approaching zero-temperature metallic states in mesoscopic superconductor--normal--superconductor arrays},
	url = {https://doi.org/10.1038/nphys2154},
	volume = {8},
	year = {2012}}

@article{ dalidovich_nonlinear_2004,
  title = {Nonlinear Transport near a Quantum Phase Transition in Two Dimensions},
  author = {Dalidovich, Denis and Phillips, Philip},
  journal = {Phys. Rev. Lett.},
  volume = {93},
  issue = {2},
  pages = {027004},
  numpages = {4},
  year = {2004},
  month = {Jul},
  publisher = {American Physical Society},
  doi = {10.1103/PhysRevLett.93.027004},
  url = {https://link.aps.org/doi/10.1103/PhysRevLett.93.027004}
}

@article{ green_nonlinear_2005,
  title = {Nonlinear Quantum Critical Transport and the Schwinger Mechanism for a Superfluid-Mott-Insulator Transition of Bosons},
  author = {Green, A. G. and Sondhi, S. L.},
  journal = {Phys. Rev. Lett.},
  volume = {95},
  issue = {26},
  pages = {267001},
  numpages = {4},
  year = {2005},
  month = {Dec},
  publisher = {American Physical Society},
  doi = {10.1103/PhysRevLett.95.267001},
  url = {https://link.aps.org/doi/10.1103/PhysRevLett.95.267001}
}

@article{vigneau_probing_2023,
    author = {Vigneau, Florian and Fedele, Federico and Chatterjee, Anasua and Reilly, David and Kuemmeth, Ferdinand and Gonzalez-Zalba, M. Fernando and Laird, Edward and Ares, Natalia},
    title = {Probing quantum devices with radio-frequency reflectometry},
    journal = {Applied Physics Reviews},
    volume = {10},
    number = {2},
    pages = {021305},
    year = {2023},
    month = {02},
    issn = {1931-9401},
    doi = {10.1063/5.0088229},
    url = {https://doi.org/10.1063/5.0088229},
    eprint = {https://pubs.aip.org/aip/apr/article-pdf/doi/10.1063/5.0088229/18145889/021305\_1\_5.0088229.pdf},
}

@article{barbara_stimulated_1999,
  title = {Stimulated Emission and Amplification in Josephson Junction Arrays},
  author = {Barbara, P. and Cawthorne, A. B. and Shitov, S. V. and Lobb, C. J.},
  journal = {Phys. Rev. Lett.},
  volume = {82},
  issue = {9},
  pages = {1963--1966},
  numpages = {0},
  year = {1999},
  month = {Mar},
  publisher = {American Physical Society},
  doi = {10.1103/PhysRevLett.82.1963},
  url = {https://link.aps.org/doi/10.1103/PhysRevLett.82.1963}
}

@article{cawthorne_synchronized_1999,
  title = {Synchronized oscillations in Josephson junction arrays: The role of distributed coupling},
  author = {Cawthorne, A. B. and Barbara, P. and Shitov, S. V. and Lobb, C. J. and Wiesenfeld, K. and Zangwill, A.},
  journal = {Phys. Rev. B},
  volume = {60},
  issue = {10},
  pages = {7575--7578},
  numpages = {0},
  year = {1999},
  month = {Sep},
  publisher = {American Physical Society},
  doi = {10.1103/PhysRevB.60.7575},
  url = {https://link.aps.org/doi/10.1103/PhysRevB.60.7575}
}

@Article{charpentier_first_2025,
  author={Charpentier, Thibault
  and Perconte, David
  and L{\'e}ger, S{\'e}bastien
  and Amin, Kazi Rafsanjani
  and Blondelle, Florent
  and Gay, Fr{\'e}d{\'e}ric
  and Buisson, Olivier
  and Ioffe, Lev
  and Khvalyuk, Anton
  and Poboiko, Igor
  and Feigel'man, Mikhail
  and Roch, Nicolas
  and Sac{\'e}p{\'e}, Benjamin},
  title={First-order quantum breakdown of superconductivity in an amorphous superconductor},
  journal={Nature Physics},
  year={2025},
  month={Jan},
  day={01},
  volume={21},
  number={1},
  pages={104-109},
  issn={1745-2481},
  doi={10.1038/s41567-024-02713-8},
  url={https://doi.org/10.1038/s41567-024-02713-8}
}

@article{weitzel_sharpness_2023,
  title = {Sharpness of the Berezinskii-Kosterlitz-Thouless Transition in Disordered NbN Films},
  author = {Weitzel, Alexander and Pfaffinger, Lea and Maccari, Ilaria and Kronfeldner, Klaus and Huber, Thomas and Fuchs, Lorenz and Mallord, James and Linzen, Sven and Il'ichev, Evgeni and Paradiso, Nicola and Strunk, Christoph},
  journal = {Phys. Rev. Lett.},
  volume = {131},
  issue = {18},
  pages = {186002},
  numpages = {6},
  year = {2023},
  month = {Nov},
  publisher = {American Physical Society},
  doi = {10.1103/PhysRevLett.131.186002},
  url = {https://link.aps.org/doi/10.1103/PhysRevLett.131.186002}
}

@article{schoelkopf_rfset_1998,
  author = {R. J. Schoelkopf  and P. Wahlgren  and A. A. Kozhevnikov  and P. Delsing  and D. E. Prober },
  title = {The Radio-Frequency Single-Electron Transistor (RF-SET): A Fast and Ultrasensitive Electrometer},
  journal = {Science},
  volume = {280},
  number = {5367},
  pages = {1238-1242},
  year = {1998},
  doi = {10.1126/science.280.5367.1238}
}

@article{gantmakher_superconductor_2010,
doi = {10.3367/UFNe.0180.201001a.0003},
url = {https://dx.doi.org/10.3367/UFNe.0180.201001a.0003},
year = {2010},
month = {jan},
publisher = {},
volume = {53},
number = {1},
pages = {1},
author = {Gantmakher, Vsevolod F and Dolgopolov, Valery T},
title = {Superconductor–insulator quantum phase transition},
journal = {Physics-Uspekhi}
}

\onecolumngrid
\clearpage
\section*{\SM} 

\renewcommand{\thesection}{SM~\Roman{section}}
\renewcommand{\thesubsection}{\alph{subsection}}
\setcounter{section}{0}
\renewcommand{\thefigure}{S\arabic{figure}}
\setcounter{figure}{0}

\begin{figure}[!ht]
	\centering\includegraphics[width=6.55in]{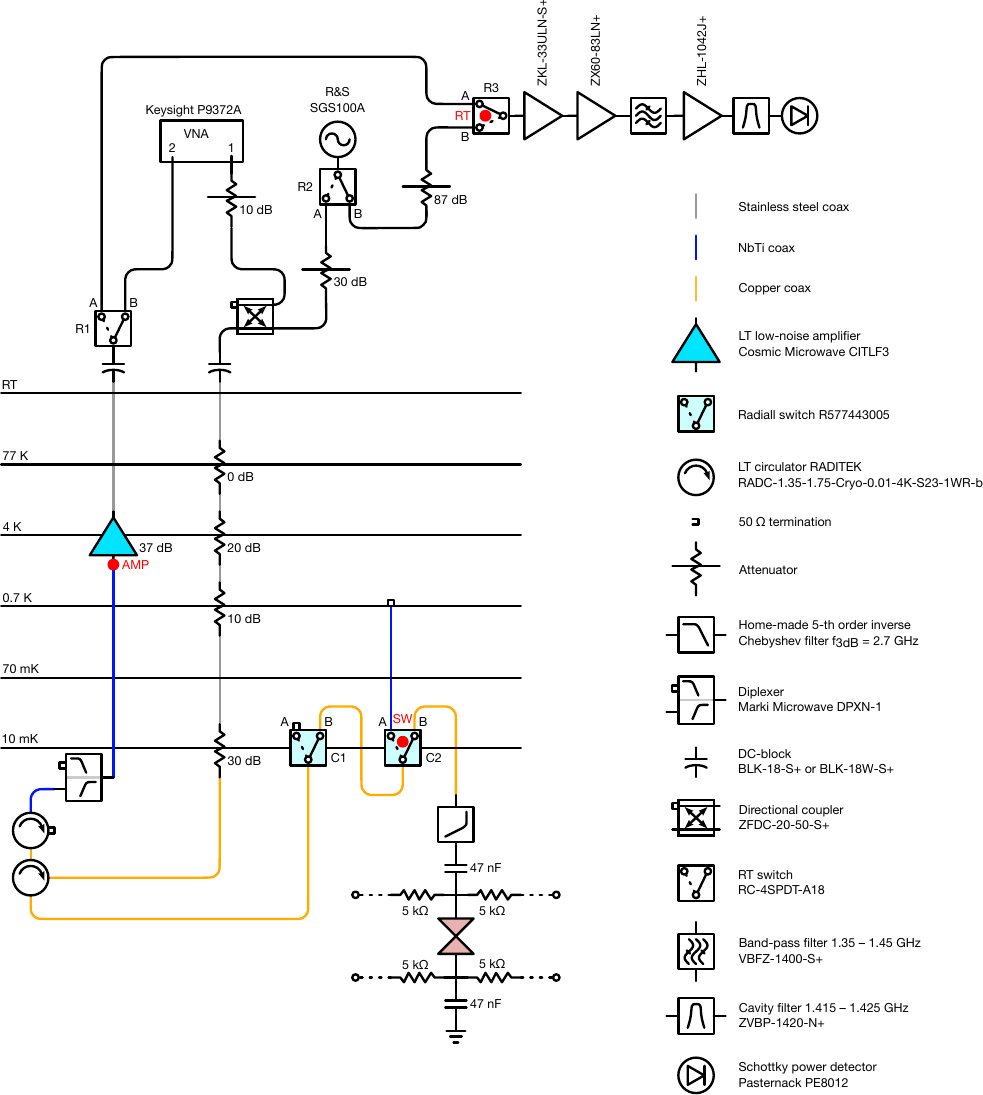}
	\caption{\textbf{Experimental setup used in this work (MW part).} 
	All elements located below the horizontal line designated with `RT' are installed within the dilution refrigerator. 
	Microwave signals from the VNA are attenuated by $10~\mathrm{dB}$ and then coupled to the sample via RF directional coupler with $20~\mathrm{dB}$ of insertion loss. 
	Microwave signals from a signal generator are sent through $30~\mathrm{dB}$ of attenuation before entering the fridge. 
	A highly attenuated line leading to a circulator on the $10~\mathrm{mK}$ stage of the fridge brings microwave signals to the sample. 
	The sample is shown on the lower right-hand side by a circuit schematic. 
	Two switches (C1, C2) are installed on the $10~\mathrm{mK}$ stage which allow radiation from either the sample, a $50~\mathrm{\Omega}$ cap on the $700~\mathrm{mK}$ stage, or from a $50~\mathrm{\Omega}$ cap on the $10~\mathrm{mK}$ stage to be transmitted to the low-temperature amplifier on the $4~\mathrm{K}$ stage. 
	The low-temperature amplifier has a noise temperature of approximately $4~\mathrm{K}$ and gain of about $35~\mathrm{dB}$ at $1~\mathrm{GHz}$. 
	Radiation is amplified at $4~\mathrm{K}$ stage and then transmitted to the room temperature amplification chain. 
	Noise is measured with a power diode.
	\vspace{-30pt}} % to force figure stay on this page
	\label{fig:fig_fridge}
\end{figure}

\begin{figure}[!ht]
	\centering\includegraphics[width=7.20in]{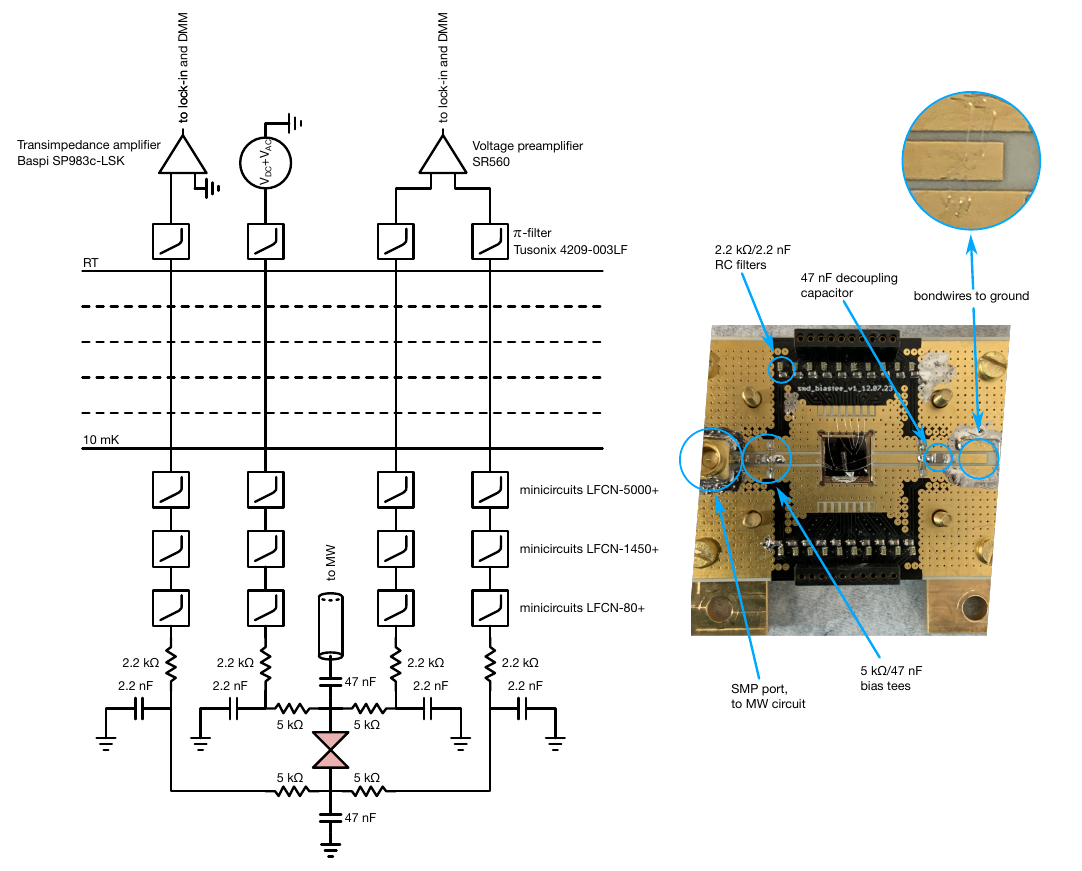}
	\caption{\textbf{Experimental setup used in this work (DC part).} 
  DC and lock-in measurements in this work were performed using the setup shown here.
  The TU Delft IVVI rack was used to source DC, combine it with the lock-in generator output and apply it to the sample.
  All lines are heavily filtered at both room temperature and cryogenic stages, and cryogenic wiring is realized with constantan twisted-pair looms.
  On the right side optical micrograph picture of the device mounted on a sample holder is shown.}
	\label{fig:fig_fridge_dc}
\end{figure}

\section{Calibration}
\label{sec:calibration}
\reffig{fig:fig_fridge} shows the setup of components installed at room temperature and inside the dilution refrigerator used in this work. 
We perform two types of calibration procedures.
The first is to rule out the weak drift of the room temperature (RT) amplifiers.
In \reffig[a]{fig:responsivity}, noise power as measured by the power diode (in volts) is plotted as a function of time, while the nominal power on the input of the room temperature chain is constant.
We observe two timescales of drift, the first being fast oscillations with a period of about 30 minutes.
The second timescale is a slow drift of the envelope of the fast oscillations on a timescale of tens of hours.
The amplitude of the fast oscillations is about $15~\mathrm{mK}$ when converted to temperature units.
We attribute these effects either to temperature variations in the laboratory room or to a slow drift of the various power supplies or other equipment.

To effectively remove these drifts, we measure a responsivity curve of the RT chain at each data point.
We use the signal generator with a frequency of $1.42~\mathrm{GHz}$ (center of the cavity filter band) to sweep over a range of output powers that cover the range of the noise powers (in volts) irradiated from the fridge (switch `RT' is in position `B' in \reffig{fig:fig_fridge}).
Then we flip the switch to position `A' and measure noise power emitted from the fridge through the same chain as the signal generator power.
This makes it possible to match an output voltage of the power diode to a corresponding power in watts at the point `RT' in the diagram.
Ideally, this removes the fluctuations due to a drift in the room temperature amplification chain, as shown in \reffig[b]{fig:responsivity}.
The blue curve is the same voltage shown in panel a, but it has been converted to power units by a responsivity curve taken at time $t=0$.
The orange curve is the same data, but calibrated at each data point by a responsivity curve taken at the same time.
The long term drift has been effectively eliminated, and the short term oscillations are greatly reduced. 
The amplitude of oscillations of the orange trace is about $4~\mathrm{mK}$ in temperature units.

\begin{figure}[!ht]
	\centering\includegraphics[width=5.5in]{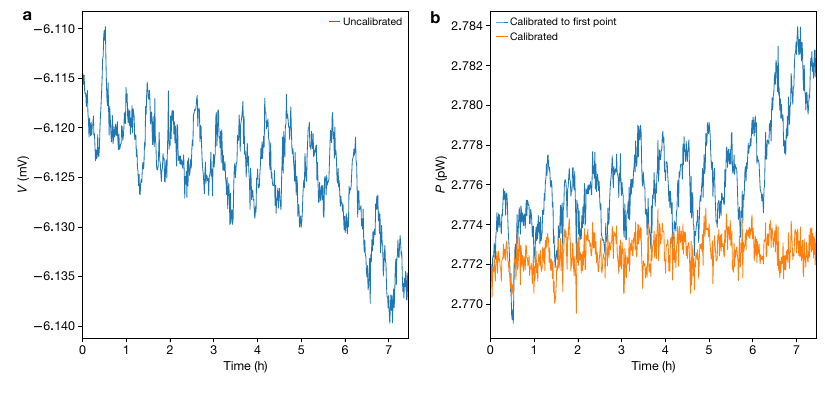}
	\caption{\textbf{Responsivity curve calibration.} 
	\textbf{a},~The drift in measured noise on the power diode as a function of time. 
	\textbf{b},~The effect of performing a responsivity calibration to reduce drifts.
	The blue trace is the same data shown in panel a, but using a responsivity curve taken at the first point to calibrate the whole trace.
	The orange trace is the same data but with a responsivity curve taken at each time point and used to calibrate each point.}
	\label{fig:responsivity}
\end{figure}

The second calibration procedure we apply is done to convert the noise power at `RT' measured in watts into temperature units at point `SW', see diagram in \reffig{fig:fig_fridge}.
To do this, we measure the radiation emanating from the sample (at 0 gate voltage), from the $50~\mathrm{\Omega}$ cap on the $700~\mathrm{mK}$ `still' stage, and from the $50~\mathrm{\Omega}$ cap on the mixing chamber stage as a function of temperature from base ($\sim 50~\mathrm{mK}$) to $1~\mathrm{K}$.

The results of these measurements are shown in \reffig{fig:fig_calibration}. 
In the left panel, the measured noise power (referred to the point `RT' as described above) is plotted as a function of the temperature of the stage where the sample or $50~\mathrm{\Omega}$ cap is located, i.e. the sample and $50~\mathrm{\Omega}$ cap on the mixing chamber stage are plotted versus the temperature of the mixing chamber stage ($T_\mathrm{cryo}$) as it is swept from base temperature to $1~\mathrm{K}$ and the noise from the $50~\mathrm{\Omega}$ cap on the still stage is plotted versus the temperature of the still stage as the temperature of the mixing chamber is swept from base to $1~\mathrm{K}$.

In the right panel, all measured noise is plotted as a function of mixing chamber ($T_\mathrm{cryo}$).
Although it is intuitive that noise power emitted from the load should be proportional to the load temperature, we see in \reffig[a]{fig:fig_calibration} that this is not the case for the load located on the still stage.
This discrepancy can be eliminated if a loss, $\eta$, between the `SW' and `AMP' points is introduced.
This loss couples irradiation from the load on the still plate with the irradiation from the bath with $T_\mathrm{cryo}$, where we implied that the bath temperature is at $T_\mathrm{cryo}$ because all the lossy components are installed on the mixing chamber plate.
Thus, the suggestive model for the measured power emitted from loads located at different stages is:
\begin{align}
    \label{cal_still}
    P_\mathrm{still}/k_B &= G (1 - \eta) T_\mathrm{still} + G \eta T_\mathrm{cryo} + G T_\mathrm{add} \\
    \label{cal_mxc}
    P_\mathrm{cryo}/k_B &= G (T_\mathrm{cryo} + T_\mathrm{add})
\end{align}
where $P_\mathrm{still}$ is the noise power measured at the point `RT' coming from the $50~\mathrm{\Omega}$ cap on the still stage, $G$ is the gain of the low temperature amplifier, $\eta$ is the loss between the points `SW' and `AMP', $T_\mathrm{still}$ is the temperature of the still stage, $T_\mathrm{cryo}$ is the temperature of the mixing chamber stage, $T_\mathrm{add}$ is the added noise temperature of low temperature amplifier, and $P_\mathrm{cryo}$ is the noise measured at `RT' coming from the $50~\mathrm{\Omega}$ cap on the mixing chamber stage or from the sample (for a sample which is in equilibrium with the fridge). 

The gain $G$ and added noise temperature $T_\mathrm{add}$ of the low temperature amplifier are extracted from the slope and intercept of the linear fit to \refeq{cal_mxc}. 
The insertion loss $\eta$ is extracted from the least-squares fit to \refeq{cal_still}, using the $G$ and $T_\mathrm{add}$ found from fitting to \refeq{cal_mxc}.
The values extracted from these fits are shown in the legend of \reffig{fig:fig_calibration}.
We repeat this calibration procedure for each cooldown and find similar but slightly different values for each fitting parameter, and use those values for calculating the excess noise in temperature units $P/k_B$, and the sample temperatures $T_\mathrm{s}$ reported in the main text. 
The values we obtain fall within manufacturer specifications.

\begin{figure}
	\centering\includegraphics[width=5.50in]{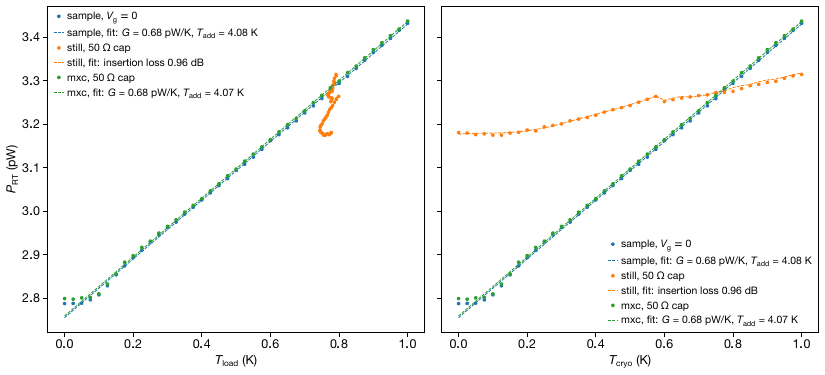}
	\caption{\textbf{Details on chain calibration.} 
	We calibrate the gain $G$ and added noise temperature $T_\mathrm{add}$ of LTA, and insertion loss $\eta$.
	The left panel shows the noise measured from the sample, the $50~\mathrm{\Omega}$ cap on the $700~\mathrm{mK}$ (`still') stage, and the $50~\mathrm{\Omega}$ cap on the mixing chamber stage as a function of the temperature of their stage, $T_\mathrm{load}$.
	The right panel shows the same quantities but plotted as a function of the temperature of the mixing chamber stage, $T_\mathrm{cryo}$.
	The gain found by linear fit to the sample and $50~\mathrm{\Omega}$ cap on the mixing chamber stage data corresponds to $G=36.9~\mathrm{dB}$ (for $10~\mathrm{MHz}$ measurement band) and is in good agreement with manufacturer specs.}
	\label{fig:fig_calibration}
\end{figure}

\subsection{Calibration in Fig.~\ref{fig:fig4}}
\label{sec:cal_sn}
In Fig.~\ref{fig:fig4} we introduced a new symbol $\widetilde{T}_\mathrm{s}$, which is conceptually the same as $T_\mathrm{s}$ but uses slightly different system parameters.
Explicitly, $\widetilde{T}_\mathrm{s}$ is derived from the same relation as Eq.~\ref{eq:beamsplitter}
\begin{equation}
P/k_B = -\alpha |S_{11}|^2~(\widetilde{T}_\mathrm{s} - \widetilde{T}_\mathrm{cryo}) + \widetilde{T}_\mathrm{s}.
\end{equation}
The distinction is that to determine $\widetilde{T}_\mathrm{s}$ we fix $\widetilde{T}_\mathrm{cryo}$ from the measured noise power $P$ at a gate voltage where $\alpha |S_{11}|^2 \approx 1$.
In contrast, to determine $T_\mathrm{s}$ we simply fixed $T_\mathrm{cryo}=50~\mathrm{mK}$.
$\widetilde{T}_\mathrm{s}$ is useful for datasets containing gate-voltage sweeps because it serves to cancel out long-time drifts.
We also note that the data in Fig.~\ref{fig:fig4} use $\alpha=64.5~\mathrm{dB}$, fit from a different cooldown (Fig.~\ref{fig:fig3}).
We have verified that plausible differences in $\alpha$ do not affect the collapse in Fig.~\ref{fig:fig4}.

\section{Circuit theory}
\label{sec:circuit_theory}

\reffig[a]{fig:schematic} shows a schematic of a simplified model circuit for our measurement.
A sample modeled as a voltage generator $V_\mathrm{s}$ with output impedance $Z_\mathrm{s}$ is connected to a transmission line of characteristic impedance $Z_0$ and propagation constant $\beta$.
The transmission line has a left-moving scattering state $V^{-}(x) = V^{-} e^{\mathrm{i} \beta x}$ and right-moving scattering state $V^{+}(x) = V^{+} e^{- \mathrm{i} \beta x}$.
An ideal circulator followed by an impedance-matched amplifier is used to measure $V^{+}(L)$.
The third port of the circulator is terminated by a matched load $Z_0$, with an effective Johnson-Nyquist voltage generator $V_\mathrm{c}$.

\begin{figure}[h]
	\centering\includegraphics[width=3.8in]{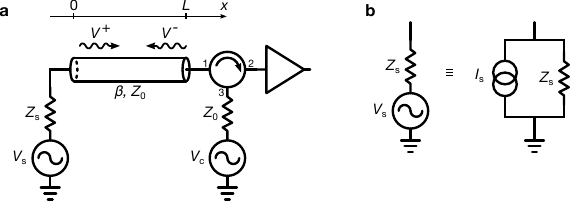}
	\caption{\textbf{Circuit model schematic.}
	\textbf{a},~Voltage generator model for our circuit.
	\textbf{b},~Equivalent current generator circuit.}
	\label{fig:schematic}
\end{figure}

Applying Ohm's law to the third port of the circulator yields 
\begin{align}
	V^{-}(L) = V_\mathrm{c}/2,
	\label{eq:vl2vc}
\end{align}
as one might expect for a generator with a matched load.
Current conservation for the sample connected to a transmission line reads:
\begin{align}
    \frac{V_\mathrm{s} - (V^{+}(0) + V^{-}(0))}{Z_\mathrm{s}} = \frac{V^{+}(0) - V^{-}(0)}{Z_0},
    \label{eq:current_conservation}
\end{align}
where we have used the fact that the current through the transmission line $I(x)=(V^{+}(x) - V^{-}(x)) / Z_0$ with the positive direction of the current being towards the right.
Rearranging \refeq{eq:current_conservation} gives
\begin{align*}
	V^{+}(0) = V^{-}(0) \frac{Z_\mathrm{s} - Z_0}{Z_\mathrm{s} + Z_0} + V_\mathrm{s} \frac{Z_0}{Z_\mathrm{s} + Z_0}.
\end{align*}
Finally, rewriting this equation at $x=L$ and substituting \refeq{eq:vl2vc} we obtain
\begin{align*}
    V^{+}(L) = \frac{V_\mathrm{c}}{2} \frac{Z_\mathrm{s} - Z_0}{Z_\mathrm{s} + Z_0} e^{-2 \mathrm{i} \beta L} + V_\mathrm{s} \frac{Z_0}{Z_\mathrm{s} + Z_0} e^{-\mathrm{i} \beta L}.
\end{align*}
The first term on the right-hand side is the reflected wave generated by the load of the circulator, and the second is the wave generated by the sample's voltage generator.
The voltage spectral density seen by the amplifier measured in the frequency band $\Delta f$ is $S_{V}^{+}=\langle V^{+}(L)^2\rangle/\Delta f$, which can be expressed as
\begin{align}
	S_{V}^{+} = \frac{S_{V}^\mathrm{c}}{4} \left|\frac{Z_\mathrm{s} - Z_0}{Z_\mathrm{s} + Z_0}\right|^2 + S_{V}^\mathrm{s} \left|\frac{Z_0}{Z_\mathrm{s} + Z_0}\right|^2,
	\label{eq:srr}
\end{align}
where we have used the fact that $V_\mathrm{c}$ and $V_\mathrm{s}$ are uncorrelated.
The Johnson-Nyquist sample temperature is defined through $S_{V}^\mathrm{s} = 4 k_B T_\mathrm{s} \Re[Z_\mathrm{s}]$ and the circulator temperature through $S_{V}^\mathrm{c} = 4 k_B T_\mathrm{c} Z_0$.
These quantities are related to the measured input noise power spectral density $P = S_{V}^{+}/Z_0$ by
\begin{equation}
  P = k_B T_\mathrm{c} |S_{11}|^2 + k_B T_\mathrm{s} (1-|S_{11}|^2),
\end{equation}
where the reflection coefficient for the sample $S_{11}$ can be expressed as usual via the sample impedance $Z_\mathrm{s}$ as $S_{11} = \frac{Z_\mathrm{s} - Z_0}{Z_\mathrm{s} + Z_0}$. Incorporating the net loss $\alpha$ gives \refeq{eq:beamsplitter} in the main text.
Note that the result depends only on the magnitude of reflection $|S_{11}|$, so a phase calibration is not needed.

Alternatively, the sample noise can also be expressed as arising from an equivalent current generator, as indicated in \reffig[b]{fig:schematic}.
The current noise spectral density $S_I$ is related to the sample voltage noise spectral density by $S_I^\mathrm{s} = S_V^\mathrm{s}/|Z_\mathrm{s}|^2$.
Substituting into \refeq{eq:srr} and using the same temperature definitions results in
\begin{equation}
  P = k_B T_\mathrm{c} |S_{11}|^2 + S_I^\mathrm{s} Z_0 \frac{|Z_\mathrm{s}|^2}{|Z_\mathrm{s} + Z_0|^2}.
\end{equation}
We see that for samples with real impedance $Z_\mathrm{s}$ when $Z_\mathrm{s} \gg Z_0$ the measured noise power spectral density directly gives $S_I^\mathrm{s}$, $P \approx k_B T_\mathrm{c} + S_I^\mathrm{s} Z_0$.
However in the general case, both the magnitude and phase of $S_{11}$ must be known
\begin{equation}
  P = k_B T_\mathrm{c} |S_{11}|^2 + \frac{1}{4} S_I^\mathrm{s} Z_0 |1 + S_{11}|^2.
\end{equation}
Fano factors are estimated from the measured data using the above equation.
The phase of $S_{11}$ is estimated based on a comparison with a distributed $RC$ model in the normal state.

\section{Inductance estimates}
\label{sec:inductance_estimates}
In the previous section we showed how the experimentally measurable reflection coefficient $S_{11}$ can be used to translate between output noise power spectral density and device temperature.
The relations we derived do not depend on microscopic details of the device.

Although not required for temperature inference, it is still interesting to understand the behavior of the measured reflection coefficient in more detail.
% In addition to the above derivation, we have also performed extensive analysis of a transmission line model focusing on the device itself. 
We have found that the device can be approximated as a one-dimensional transmission line, as shown in \reffig{fig:sample_impedance}. 
We extend this model further to include a term accounting for a gate input impedance ($Z_\mathrm{g,in}$), and a series impedance associated with the bonding pads and wires ($Z_\mathrm{in}$), and scaling factors to account for offsets in the magnitude and phase of the measured $|S_{11}|$.
% Using least squares regression techniques we are able to extract reasonable values for the Josephson inductance across the SIT.
% We additionally compare the circuit model-based estimates of inductance to several other methods and find good agreement, supporting the legitimacy of our model.

\begin{figure}
  \centering\includegraphics[width=4.0in]{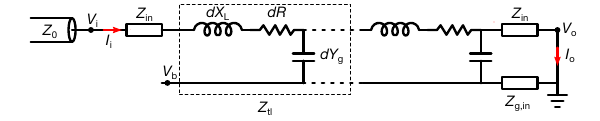}
  \caption{\textbf{Sample impedance calculation.}
  Device circuit model, consisting of a transmission line with impedance $Z_\mathrm{tl}$ and input impedances $Z_\mathrm{in}$, $Z_\mathrm{g,in}$.
  A transmission line with characteristic impedance $Z_\mathrm{tl}$ to inductive reactance per unit length $d X_L$, capacitive susceptance to ground $d Y_g$, and resistance per unit length $d R$ is connected via line input impedance $Z_\mathrm{in}$ and gate input impedance $Z_\mathrm{g,in}$.
  }
  \label{fig:sample_impedance}
\end{figure}

Here we derive the continuum expression for fitting to the device model. 
We have numerically checked that a discrete-model expression gives the same results.
We start with the ABCD matrix-expression for a transmission line
\begin{align}
  \label{eq:inout}
  \begin{bmatrix}
  V_\mathrm{i} - V_\mathrm{b} \\
  I_\mathrm{i} 
  \end{bmatrix} = T
  \begin{bmatrix}
    V_\mathrm{o} - V_\mathrm{b} \\
    I_\mathrm{o} 
  \end{bmatrix},
\end{align}
where $V_\mathrm{i} - V_\mathrm{b}$ is the potential difference at the input port, $V_\mathrm{o} - V_\mathrm{b}$ is the potential difference at the output port, $I_\mathrm{i}$ and $I_\mathrm{o}$ are the currents flowing into and out of the input and output ports. $V_\mathrm{b}$ is thus defined as $V_\mathrm{b} = (I_\mathrm{i} - I_\mathrm{o})Z_\mathrm{g,in}$, for $Z_\mathrm{g,in}$ series impedance associated with the top-gate shown in \reffig{fig:sample_impedance}. $T$ is the transmission matrix given by
\begin{align*}
  T = \begin{bmatrix}
        \cosh \gamma & Z_\mathrm{tl} \sinh \gamma \\
        \frac{1}{Z_\mathrm{tl}}\sinh \gamma & \cosh \gamma 
  \end{bmatrix}
  \begin{bmatrix}
    1 & Z_\mathrm{in} \\
    0 & 1
  \end{bmatrix},
\end{align*}
where $Z_\mathrm{in}$ is the output-side series impedance associated with bonding pads and wires indicated in \reffig{fig:sample_impedance}, $Z_\mathrm{tl} = \sqrt{\frac{\mathrm{d}R + i \mathrm{d}X_\mathrm{L}}{i \mathrm{d}Y_\mathrm{g}}}$.
As in \reffig{fig:sample_impedance}, $V_\mathrm{o}$ is grounded so we can rearrange \refeq{eq:inout} to find:
\begin{align*}
  \begin{bmatrix}
    V_\mathrm{i} \\
    I_\mathrm{i}
  \end{bmatrix}
  = T \begin{bmatrix}
    -V_\mathrm{b} \\
    I_\mathrm{o}
  \end{bmatrix} +
  \begin{bmatrix}
    V_\mathrm{b} \\
    0
  \end{bmatrix} = \left( T +
  \begin{bmatrix}
     -1 & 0 \\
     0 & 0
  \end{bmatrix} \right)
  \begin{bmatrix}
    -V_\mathrm{b} \\
    I_\mathrm{o}
  \end{bmatrix}
\end{align*}
defining $T'$ as the matrix preceding $\big[\begin{smallmatrix} 
-V_\mathrm{b} \\
I_\mathrm{o} \end{smallmatrix}\big]$ in the above equation. 
This can then be rewritten as
\begin{align*}
  \begin{bmatrix}
    V_\mathrm{i} \\
    I_\mathrm{i}
  \end{bmatrix} = 
  T' \begin{bmatrix}
    I_\mathrm{o}Z_\mathrm{g,in} \\
    I_\mathrm{o}
  \end{bmatrix} - T'
  \begin{bmatrix}
    I_\mathrm{i}Z_\mathrm{g,in} \\
    0
  \end{bmatrix}
\end{align*}
and then further simplified as
\begin{align*}
  T' \begin{bmatrix}
    Z_\mathrm{g,in} \\
    1
  \end{bmatrix} I_\mathrm{o}
  = \begin{bmatrix}
    V_\mathrm{i} \\
    I_\mathrm{i}
  \end{bmatrix}
  + T' \begin{bmatrix}
    I_\mathrm{i}Z_\mathrm{g,in} \\
    0
  \end{bmatrix}
\end{align*}
out of which we can define $T''$ and come to the final expression
\begin{align*}
  \begin{bmatrix}
    V_\mathrm{i} \\
    I_\mathrm{i}
  \end{bmatrix} =
  \big(T''\big)^{-1}T' \begin{bmatrix}
    Z_\mathrm{g,in} \\
    1
  \end{bmatrix} I_\mathrm{o}
\end{align*}
which is analytically solvable, and whose solution for the device impedance is
\begin{align}
  \label{eq:inputimped}
  Z_\mathrm{dev} = \frac{\left(Z_\mathrm{in}+ 2Z_\mathrm{g,in} \right)Z_\mathrm{tl} \cosh \gamma + \left( Z_\mathrm{in} Z_\mathrm{g,in} + Z_\mathrm{tl}^2 \right) \sinh \gamma - 2Z_\mathrm{g,in}Z_\mathrm{tl}}{Z_\mathrm{tl} \cosh \gamma + \left(Z_\mathrm{in} + Z_\mathrm{g,in} \right) \sinh \gamma}
\end{align}
we then use this \refeq{eq:inputimped} to estimate the total impedance of our sample $Z_\mathrm{s} = Z_\mathrm{dev} + Z_\mathrm{in}$. As a check we can substitute $Z_\mathrm{g,in} = 0$ into the above expression to recover the formula for a terminated transmission line
\begin{align*}
  Z_\mathrm{dev} = Z_\mathrm{tl} \frac{Z_\mathrm{in} + Z_\mathrm{tl} \tanh \gamma}{Z_\mathrm{tl} + Z_\mathrm{in} \tanh \gamma}.
\end{align*}
% otherwise known as the lossy telegrapher's equation.

While the general equation \refeq{eq:inputimped} can be used to fit our data with great precision, we have found that it is sufficient to assume a fully real input impedance $Z_\mathrm{in} = R_\mathrm{in}$, dominated by the resistance of the aluminum bond pads in the normal state, and to neglect the input gate impedance  $Z_\mathrm{g,in}$, simplifying the fitting procedure considerably.
This reduces the number of free fitting parameters in the normal state (above $T_\mathrm{c}$) to just 4 parameters. 

% \begin{figure}[!ht]
% 	\centering\includegraphics[width=3.54in]{figS_Lest_3.54in_ink.pdf}
% 	\caption{\textbf{Gate dependence of S-parameters and fits.}
% 	The real and imaginary components of the S-parameters are plotted versus gate voltage along with the results of the fitting procedure. 
% 	In panels \textbf{a} and \textbf{b} the sample is measured at approximately $1.5~\mathrm{K}$, well above the critical temperature, and the fit was performed with 4 free parameters. 
% 	In panels \textbf{c} and \textbf{d} the sample is measured while the cryostat is held at $50~\mathrm{mK}$, and the fit was performed for each gate voltage with a single free parameter.}
% 	\label{fig:s11_fits}
% \end{figure}

The procedure we use to estimate the inductance of the sample is as follows. 
In the normal state, i.e. at $1.5~\mathrm{K}$, we analyze a trace taken over a range of gate voltages which span the SIT at $50~\mathrm{mK}$.
Using measured device resistance ($R$), we fit the measured $S_{11}$ to \refeq{eq:inputimped} (transformed using $S_{11} = \frac{Z_\mathrm{s} - Z_0}{Z_\mathrm{s} + Z_0}$).
We allow 4 parameters to vary: the real part of the input impedance $Z_\mathrm{in} = R_\mathrm{in}$, the imaginary part of the gate admittance $Y_\mathrm{g}$, and the real and imaginary parts of a complex scaling factor to account for magnitude and phase offsets in our measurement setup.
In the normal state we set the Josephson inductance ($X_\mathrm{L}$) to zero.

With this minimal set of fitting parameters we are able to fit the observed data very well, see \reffig[b\,-\,c]{fig:figS_fits_and_inductance}.
For a thin aluminum film we expect a normal state resistance $Z_\mathrm{in}$ of tens of~$\Omega$, and we obtain a best fit value of $116~\Omega$.
From the geometry of the gate we would expect a capacitance of $8~\mathrm{pF}$, and our best fit value is a capacitance of $30~\mathrm{pF}$.
Lastly, the fit gives a complex scaling factor of 2\% for $|S_{11}|$.

Then, we fix the gate admittance and complex scaling parameter to the values obtained from the fit in the normal state and apply them to the data measured at $50~\mathrm{mK}$.
At $50~\mathrm{mK}$ we set the bond pad impedance $Z_\mathrm{in}=R_
\mathrm{in}$ to zero, and the Josephson inductance is non-zero, so we allow $X_\mathrm{L} = \omega L$ to become the singular fitting parameter.
To accurately capture the changing inductance across the SIT we perform the fit at each gate voltage, until the device resistance becomes too large to measure. 
The result is a fit which well reproduces the measured S-parameters in the superconducting state and provides an estimate for the array inductance across the SIT, see \reffig[d\,-\,e]{fig:figS_fits_and_inductance}.
As shown in \reffig[f]{fig:figS_fits_and_inductance}, the inductance is found to increase rapidly with increasing resistance as the sample transits the SIT, starting from a small offset value of $1~\mathrm{nH}$.
This offset could arise from a geometrical inductance of the array, or from the kinetic inductance of the superconducting islands.

The inductance increases to a maximum value of $1~\mathrm{mH}$ just before the device becomes insulating.
Notably, at the gate voltage identified in transport measurements to be the SIT, we observe no critical behavior.
Comparing the values found via our circuit model to other estimation methods we find good agreement, as shown in \reffig[f]{fig:figS_fits_and_inductance} where we compare the circuit model to estimates drawn from the critical current ($I_\mathrm{c}$) values measured via IV curves ($L_\mathrm{J}=\frac{\hbar^2}{4I_\mathrm{c}e^2}$, assuming a sinusoidal current-phase relationship).
The estimate of $L_\mathrm{J}$ drawn from critical currents for the most negative gate voltage plotted below is drawn with a dotted line, as the zero bias resistance is not precisely zero, but there is still a well-defined region of nearly-zero bias with distinct `critical current'-like features.
% Motivated by a broadband $S_{11}$ measurement of a resonance traveling with gate voltage (visible near the middle of \reffig{fig:widef}, and below in \reffig{fig:inductances}c), we also make a crude estimate of inductance from frequency dependence of the transmission line mode.
% Assuming a capacitance of $30~\mathrm{pF}$ as found in the fitting procedure we calculate the inductance $L = \frac{1}{f^2 C}$ in the range of the traveling transmission line mode.
% We find a reasonable estimate for inductance in the superconducting state.
Inductances plotted in \reffig[f]{fig:figS_fits_and_inductance} have been shifted in gate voltage to account for device hysteresis.
The traces are shifted by an amount ($V^*$) found from the distance away from the apparent SIT defined as DC resistance reaching 60~k$\Omega$, identified in \reffig[b]{fig:fig2}.
% The resonance frequencies associated with these values of inductance are much below our measurement band, when accounting for a device capacitance set by the area of the array and thickness of the dielectric layer.

% \begin{figure}[!ht]
% 	\centering\includegraphics[width=5.5in]{figS_inductance_5.50in.pdf}
% 	\caption{\textbf{Gate dependence of different estimates for inductance.}
%   \textbf{a},~Inductance is plotted as a function of shifted gate voltage.
% 	The blue trace is inductance estimated from our circuit model.
% 	The orange trace is the inductance estimate from the critical current of IV measurements, with the dashed line indicating a point with finite zero-bias resistance but with signatures of critical current.
% 	% The  green curve is inductance estimated based on an observed resonance in a wider frequency band (see text).
%   All traces have been shifted in gate voltage to account for device hysteresis.
%   \textbf{b},~Measured resistances are plotted as a function of gate voltage with colors corresponding to the inductances plotted in \textbf{a}.
%   % \textbf{c},~The wide-frequency $|S_{11}|$ measurement with the traveling resonance emphasized by a green dashed line, also shifted as in~\textbf{a}-\textbf{b}.
%   }
% 	\label{fig:inductances}
% \end{figure}

\section{Reflection temperature dependence}
\label{sec:s11_temperature_dependence}
In \reffig{fig:fig2} in the main text we show the $|S_{11}|^2$ gate dependence at base temperature. 
Here, we show that $|S_{11}|^2$ retains the same gate dependence over a large temperature range.
\reffig{fig:s11at500} shows essentially unchanged gate dependence of the main dip centered around $-8.4~\mathrm{V}$ in $|S_{11}|^2$ over the range of base to approximately $500~\mathrm{mK}$.
At higher temperatures the structure of the gate dependence begins to change significantly.
The dip around $-8.1~\mathrm{V}$ at base temperature begins to disappear before $500~\mathrm{mK}$.
Only minor changes in $|S_{11}|^2$ are observed before the temperature at the onset of resistance saturation (approximately $150~\mathrm{mK}$).

\begin{figure}[!ht]
	\centering\includegraphics[width=2.5in]{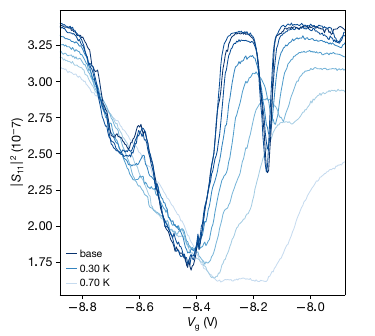}
	\caption{\textbf{Reflection gate dependence at different temperatures.}
	$|S_{11}|^2$ is plotted as a function of gate voltage for temperatures spanning from base (set point $0~\mathrm{mK}$) to $700~\mathrm{mK}$ in $100~\mathrm{mK}$ increments.
	The temperatures are indicated by colors as shown in the legend.}
	\label{fig:s11at500}
\end{figure}

\section{Finite bias noise}
\label{sec:finite_bias}
In \reffig{fig:fig4} in the main text we showed that the measured noise collapses to scaling behavior when plotted versus current, and not when plotted versus voltage.
Here, in \reffig{fig:nocollapse}, we show that this behavior is reproduced in both samples.
The dramatic collapse observed when sample temperature is plotted versus current is not reproduced when plotted versus Joule power. 
In \reffig[a,\,e]{fig:nocollapse} we plot the sample resistance as a function of current.
In \reffig[b,\,f]{fig:nocollapse} we plot the sample temperature as a function of voltage, which does not collapse as shown for device S2 in the main text.
In \reffig[c,\,g]{fig:nocollapse} we plot the output noise power spectral density $P/k_B$ as a function of current, which also does not fully collapse as sample temperature does.
In \reffig[d,\,h]{fig:nocollapse} we plot the sample temperature as a function of Joule power given by $P_\mathrm{Joule} = IV$, where the curves again do not collapse to the same extent as when plotted versus current. 

At fixed cryostat temperature we apply a large bias to the sample, heating it out of equilibrium.
Within a conventional SIT framework, sample resistance is not expected to depend on temperature at the SIT~\cite{gantmakher_superconductor_2010}.
For large enough bias, the resistance should be roughly temperature-independent.
We look for a relationship between zero-bias and high bias resistance in Fig.~\ref{fig:biasvsR}.
Plotting the ratio of zero-bias resistance to high-bias resistance as a function of gate voltage, we find unity at the zero-bias resistance identified in Fig.~\ref{fig:fig2} as the separatrix (60~k$\Omega$ for device S1).
The superconducting (`S', lilac), anomalous metallic (`AM', yellow), and insulating (`I', green) regions are indicated.
% We also investigate the criticality of the SIT by comparing the zero-bias resistance of the sample with the high-bias resistance. 
% In \reffig{fig:biasvsR} the ratio of the zero-bias to high-bias resistances as a function of gate voltage is shown for each sample.
% We look for the critical point of the SIT, identified by resistance which is temperature-independent, meaning that zero-bias and high-bias resistances are equal. \anton{why? i got an impression that we defined it as separatrix, no?}
% When plotted as the ratio of resistance at zero-bias to the resistance at high-bias, the SIT would occur around 1, marked by the dark blue line in each panel with `SIT' written vertically.
% The superconducting (S, lilac), anomalous metallic (AM, yellow), and insulating (I, green) regions are indicated in \reffig{fig:biasvsR}.
% The transition from anomalous metallic to insulating behavior marks also the SIT, as determined in \reffig[a]{fig:fig2} at the separatrix whose zero-bias resistance is about $60~\mathrm{k}\Omega$.
% This is also the point where the ratio of zero-bias to high-bias resistance is close to or equal to one. 
% The critical point of the SIT is identified by temperature-independent resistance at low temperatures, and because the application of finite bias to the sample has a heating effect, we can look for the gate voltage where the zero-bias and high-bias resistances are equal. 

\begin{figure}[!ht]
  \centering\includegraphics[width=7.0in]{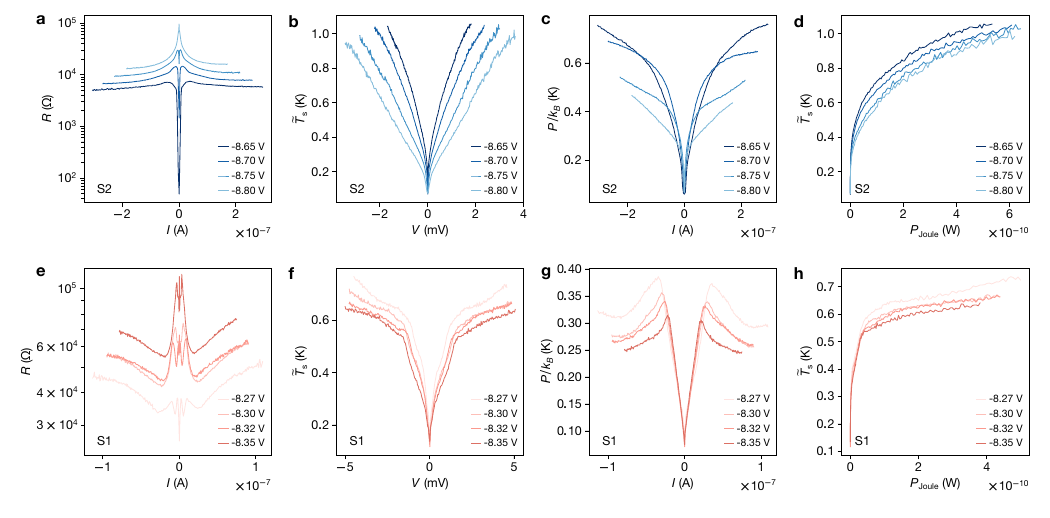}
    \caption{\textbf{Finite bias noise data.} 
    \textbf{a},~Device S1 resistance plotted as a function of current.
    \textbf{b-d},~Sample temperature plotted as a function of voltage, output noise spectral density, or power does not fully collapse. 
    \textbf{d},~Device S2 resistance plotted as a function of current.
    \textbf{e-f},~Sample temperature plotted as a function of voltage, output noise spectral density, or power does not fully collapse.}
    \label{fig:nocollapse}
\end{figure}

\begin{figure}[!ht]
	\centering\includegraphics[width=5.5in]{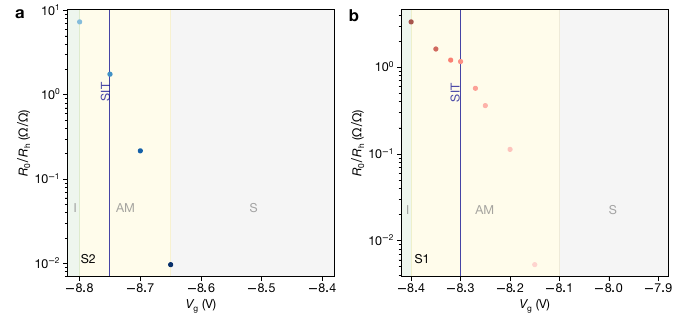}
	\caption{\textbf{Ratio between zero and high bias resistances.}
	\textbf{a},~The ratio of the resistance at zero bias, $R_0$, to the resistance at high bias, $R_\mathrm{h}$, plotted as a function of gate voltage for device S2. 
	\textbf{b},~The same ratio as a function of gate voltage for device S1. 
	In both panels regions of superconducting, S, anomalous metallic, AM, and insulating, I, behavior are indicated. 
	The regions are designated based on the zero-bias behavior of the traces.  }
	\label{fig:biasvsR}
\end{figure}

% \section{Wideband reflection measurements}
% \label{sec:wideband_reflection}
% \anton{i would just kill this paragprah and show just a figure}
% In \reffig{fig:widef} widband $|S_{11}|^2$ measured vs gate voltage for a frequency band of a circulator. 
% Overlaid in red is the DC resistance measured simultaneously, showing that the resistive state appears just after the thin dip in $|S_{11}|$ and continues throughout the wider dip region before becoming immeasurably large at the most negative gate voltages. \anton{which dip? many dips are frequency dependent}
% There is a resonance in $|S_{11}|$ around $-7.25~\mathrm{V}$, deep in the superconducting regime, which does not change with frequency.
% In contrast, the thin, pronounced dip which crosses $1.42~\mathrm{GHz}$ at about $-7.75~\mathrm{V}$, has a strong frequency dependence.
% The larger dip spanning $-7.8$ to $-8.25~\mathrm{V}$ does not have a strong frequency dependence.
% Similarly, in \reffig[b]{fig:widef} we plot the same $|S_{11}|$ as a function of gate voltage and frequency, but with the simultaneously measured $P/k_B$ overlaid in red. 
% The quantity $P/k_B$ was measured in the usual $10~\mathrm{MHz}$ band around $1.42~\mathrm{GHz}$.
% The features in $|S_{11}|$ are reproduced by the measured noise, $P/k_B$. 

% The frequency band around $1.42~\mathrm{GHz}$ was confirmed via through-line measurements without a sample to have the smoothest $|S_{11}|$ profile with the least loss from the band defined by the cryogenic circulators.

\section{Wide-frequency sweep}
\label{sec:widefreq_sweep}
We show data from a VNA sweep covering $1.1~\mathrm{GHz}$ to $1.9~\mathrm{GHz}$.
The band of the low-temperature circulators is visible between $1.2~\mathrm{GHz}$ and $1.8~\mathrm{GHz}$.
Throughout the spectrum there are standing wave ripples, and a few resonances.
Some resonances travel with decreasing gate voltage, while others appear to be independent of gate voltage.
Simple estimates of device inductance using only geometry and the resonance frequencies seen in this plot are insufficient to predict device behavior, as in the in-depth analysis in \reffig{fig:figS_fits_and_inductance}.
To compare the spectrum with other measured parameters, in \reffig[a]{fig:widef} we plot the DC resistance measured in the same gate sweep, and in \reffig[b]{fig:widef} we plot the average noise power spectral density, $P/k_\mathrm{B}$ measured in a $10~\mathrm{MHz}$ band around $1.42~\mathrm{GHz}$.

\begin{figure}[!ht]
  \centering\includegraphics[width=5.5in]{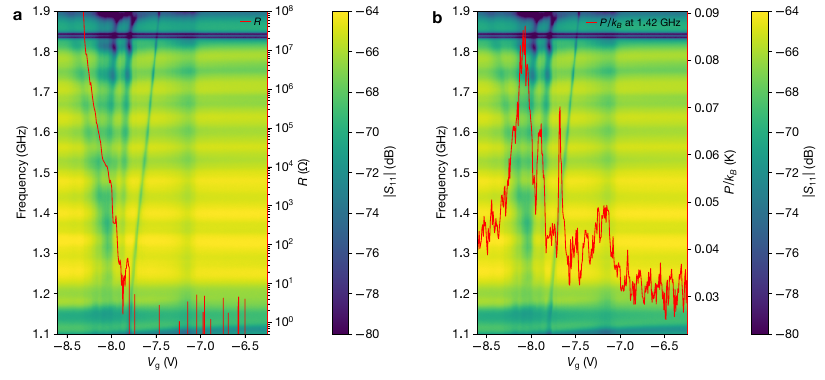}
  \caption{\textbf{Wide frequency $S_{11}$ sweep.}
  In a wide frequency band $|S_{11}|$ (color bar) is plotted as a function of frequency and gate voltage.
  \textbf{a},~Sample resistance is plotted in red on the right axis as a function of gate voltage.
  \textbf{b},~Noise power $P/k_B$ is plotted in red on the right axis as a function of gate voltage.
  Noise was measured in the same frequency band of $10~\mathrm{MHz}$ around $1.42~\mathrm{GHz}$ as in the main text}
  \label{fig:widef}
\end{figure}

\section{Noise at high bias in the normal state}
\label{sec:magnetic_field}
We apply a perpendicular magnetic field of $200~\mathrm{mT}$ to drive the sample into the normal state, testing our noise measurement technique on the normal conductor.
All measurements in \reffig{fig:bfield} are performed at zero gate voltage.
As shown in \reffig[a]{fig:bfield} on the left axis, the normal state resistance is bias independent and about $94~\Omega$.
On the right axis we observe nearly bias-independent (less than half a dB) $|S_{11}|$, about $8.5~\mathrm{dB}$ below our calibrated unity reflection of $-64.5~\mathrm{dB}$. 
In \reffig[b]{fig:bfield}, the sample temperature is plotted as a function of current, where a pronounced sub-linear behavior is observed. 
Lastly, we examine scaling behavior of the energy relaxation in the normal state.
For high bias we expect the electron-phonon relaxation to be the dominant mechanism, thus we expect $P_\mathrm{Joule}=\Sigma V T^5$, where $\Sigma$ is the electron-phonon coupling constant, $V$ is the volume, and $T$ is the temperature~\cite{giazotto_opportunities_2006}.
Fitting the sample temperature versus power $P_\mathrm{Joule}=IV$ in \reffig[c]{fig:bfield}, we find a scaling relationship above $1~\mathrm{nW}$.
The scaling exponent of $0.22$ is in good agreement with the predicted value of $1/5$. 
Having validated the electron-phonon nature of the energy relaxation we can then fit to the fixed exponent $P_\mathrm{Joule}$ vs $T_\mathrm{s}^5$ curve to find the coupling constant, $\Sigma$. 
Assuming the relevant volume of the sample is $V=7.4\times10^{-17}~\mathrm{m}^3$ (given by the sample width of $46~\mu\mathrm{m}$, length of $115~\mu\mathrm{m}$, and thickness of $14~\mathrm{nm}$), we divide the resulting slope by the volume to obtain a coupling constant $\Sigma = 3.4\times10^9~\mathrm{Wm}^{-3}\mathrm{K}^{-5}$.

\begin{figure}[!ht]
  \centering\includegraphics[width=7.0in]{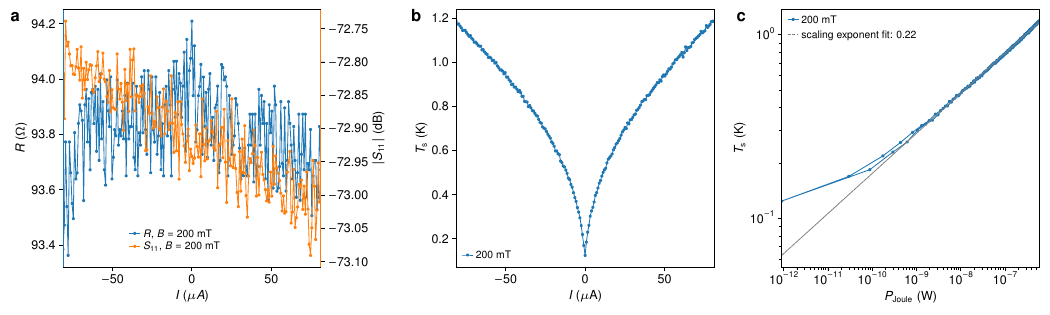}
   \caption{\textbf{Noise at high bias in normal state.}
   \textbf{a},~Resistance as a function of current plotted on the left axis in blue.
   $|S_{11}|$ as a function of current plotted on the right axis in orange.
   Resistance and $|S_{11}|$ are essentially unchanged over this bias range.
   $|S_{11}|$ is low, about $8.5~\mathrm{dB}$ below the calibrated unity reflection. 
   \textbf{b},~Sample temperature plotted as a function of current.
   The slope is higher at low bias, and lower at high bias.
   \textbf{c},~Sample temperature plotted as a function of power, $P_\mathrm{Joule}=IV$.
   At higher powers there is a scaling relationship with exponent approximately $1/5$.}
   \label{fig:bfield}
\end{figure}

\section{Additional sample}
\label{sec:additional_sample}
An additional sample device S0 was loaded to the dilution refrigerator in a different measurement configuration without the cryogenic switch (so \textit{in-situ} calibration cannot be performed).
The measurement setup had less filtering and higher noise level than the setup used in this work and presented in \reffig{fig:fig_fridge}.
This device had a slightly different four-probe measurement geometry but was fabricated on a similar Al/InAs wafer and had the same number of junctions and overall orientation as device S1 in this work.
We present data in \reffig{fig:sample0} to demonstrate another anomalous metal device with corresponding excess noise spike at the onset of resistance saturation. 
In device S0, in the less well-filtered measurement chain, we observe a massive spike in excess noise of about $450~\mathrm{mK}$, which roughly corresponds to the cryostat temperature at the onset of resistance saturation. 
The noise spike presented in \reffig{fig:sample0} was measured with the lockin amplifier unplugged, and therefore no bias applied to the sample.
A separate resistance measurement with a 50~$\mu$V excitation applied to sample and fridge lines in series (different filters than in main text) is plotted as well.
In this case the noise $P'/k_B$ was calibrated by taking a temperature sweep and performing a linear fit to extract the gain (in $W/K$), and then dividing the noise power measured by the gain to get the value in $K$. 
The added noise temperature due to the amplifier is not calibrated out.
No responsivity curve or detailed hot-cold calibration with $50~\Omega$ caps was performed.

In a separate cooldown, we measured device S0 in a similarly less well-filtered, but higher frequency measurement chain.
A similar procedure was followed to extract the noise in temperature units vs gate voltage (\reffig{fig:sample0_highfreq}).
The results of the higher frequency measurement (band of $200~\mathrm{MHz}$ centered at $5.2~\mathrm{GHz}$) are qualitatively and quantitatively similar to those of the lower frequency measurement, where an estimated noise peak of just about $400~\mathrm{mK}$ is observed, again corresponding to the cryostat temperature at the onset of resistance saturation.
Here the noise data was also taken with the lockin amplifier unplugged, so that the device was measured without DC bias.
The resistance was measured in a separate measurement with 10~$\mu$V excitation applied to sample and fridge lines in series (different filters than in main text).
These results provide robust evidence that the existence of the noise peak is a general phenomenon, and is not linked solely to a specific measurement configuration or resonance condition.
Additionally, the cryostat temperature at the onset of resistance saturation, and corresponding height of the noise peak, appears to increase for less well-filtered measurement setups.

\begin{figure}[!ht]
	\centering\includegraphics[width=5.5in]{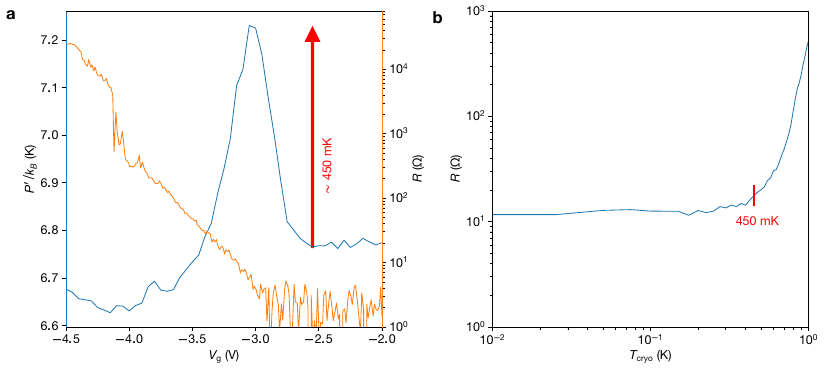}
	\caption{\textbf{Noise peak in an additional sample.}
	\textbf{a},~Noise power $P'$ of device S0 plotted in blue as a function of gate voltage.
	Most of the noise comes from the low-temperature amplifier, and due to the inadequate filtering in the measurement chain much higher noise spike is observed compared to \reffig[b]{fig:fig2}.
	Resistance is plotted as a function of gate voltage in orange on the right axis, showing that the noise spike occurs just as the device begins the SIT.
	\textbf{b},~A temperature sweep at a gate voltage of -3.2~V shows very prominent resistance saturation with an onset of around $450~\mathrm{mK}$.}
	\label{fig:sample0}
\end{figure}

\begin{figure}[!ht]
	\centering\includegraphics[width=5.5in]{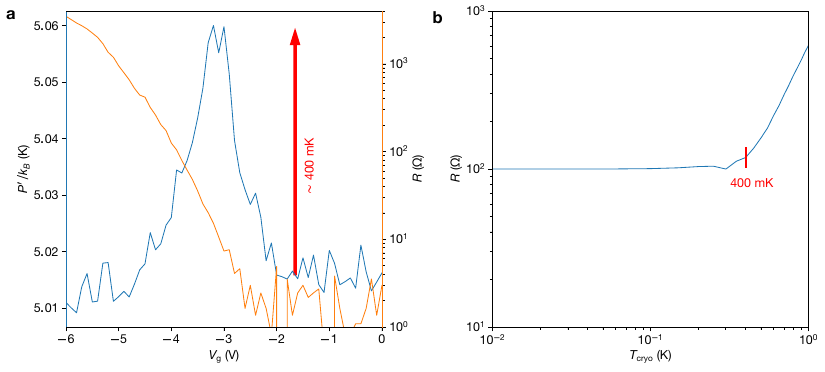}
	\caption{\textbf{Noise peak in higher frequency band.}
	\textbf{a},~Measurements of noise power $P'$ from device S0 using a spectrum analyzer with a frequency band of $200~\mathrm{MHz}$ centered at $5.2~\mathrm{GHz}$.
	A massive spike in noise is also observed (blue trace), coincident with the onset of finite resistance (orange trace, right axis).
	\textbf{b},~A temperature sweep at a gate voltage of -3.2~V shows very prominent resistance saturation with an onset of around $400~\mathrm{mK}$.
  Base temperature resistance is different in panels \textbf{a} and \textbf{b} due to hysteresis of the gate.}
	\label{fig:sample0_highfreq}
\end{figure}

\end{document}